\documentclass[twocolumn,aps,showpacs,superscriptaddress,nofootinbib,groupedaddress]{revtex4}
\usepackage{bbold}
\usepackage{graphicx}
\usepackage{amsmath}
\usepackage{amstext}
\usepackage{amsthm}
\usepackage{amssymb}
\usepackage{mathrsfs}

\newcommand{\be}{\begin{equation}}
\newcommand{\ee}{\end{equation}}
\newcommand{\ba}{\begin{eqnarray}}
\newcommand{\ea}{\end{eqnarray}}
\newcommand{\bc}{\begin{center}}
\newcommand{\ec}{\end{center}}

\newcommand{\Lc}{\Lambda_{\rm cutoff}}

\newcommand{\ka}{\hat k}
\newcommand{\kb}{\bar k}
\newcommand{\gat}{\widehat g_{\mu\nu}}
\newcommand{\bat}{\bar g_{\mu\nu}}
\newcommand{\cb}{\chi_{\rm B}}
\newcommand{\babox}{\overline{\square}}
\newcommand{\gabox}{\widehat{\square}}
\newcommand{\bet}{\boldsymbol{\beta}}

\newcommand{\gas}{\boldsymbol{g_\ast}}
\newcommand{\las}{\boldsymbol{\lambda_\ast}}

\newcommand{\tp}{\boldsymbol{\theta'}} 
\newcommand{\tpp}{\boldsymbol{\theta''}}
\newcommand{\dm}{\frac{d}{2}}
\newcommand{\SC}{\mathscr{S}}

\newcommand{\bati}{\tilde f}
\newcommand{\vo}{\mathscr{V}}

\begin{document}

\title{Universality and symmetry breaking in conformally reduced quantum gravity}
\author{Alfio Bonanno}
\email{alfio.bonanno@inaf.it}
\affiliation{
INAF -  Osservatorio Astrofisico di Catania, Via S. Sofia 78, I-95123 Catania, Italy}
\affiliation{
INFN, Via S. Sofia 64, I-95123 Catania, Italy}
\author{Filippo Guarnieri}
\email{filippo.guarnieri@aei.mpg.de}
\affiliation{Dipartimento di Fisica, Universit$\grave{a}$ degli Studi di Roma Tre and\\ 
INFN sezione di Roma Tre, Via della Vasca Navale 84, I-00146 Rome, Italy}
\affiliation{Max Planck Institute for Gravitational Physics (Albert Einstein Institute)\\
Am M\"uhlenberg 1, D-14476 Golm, Germany}

%----------------------------------------------------------------------------------
%------------------               ABSTRACT              --------------------------
%----------------------------------------------------------------------------------
\begin{abstract}
%----------------------------------------------------------------------------------
The scaling properties of quantum gravity are discussed by employing a class of proper-time regulators in the functional flow equation for the conformal factor within the formalism of the background field method. Renormalization group trajectories obtained by projecting the flow on a flat topology are more stable than those obtained from a projection on a spherical topology. In the latter case the ultraviolet flow can be characterized by a Hopf bifurcation with an ultraviolet attractive limiting cycle. Although the possibility of determining the infrared flow for an extended theory space can be severely hampered due to the conformal factor instability, we present a robust numerical approach to study the flow structure around the non-gaussian fixed point as an {inverse problem}. In particular it is shown the possibility of having a spontaneous breaking of the diffeomorphism invariance can be realized with non-local functionals of the volume operator.
\end{abstract}

\maketitle

%----------------------------------------------------------------------------------
%------------------           INTRODUCTION        --------------------------
%----------------------------------------------------------------------------------
\section{\label{sec:level1}Introduction}
%----------------------------------------------------------------------------------
In statistical mechanics universality is the property for which, close to a continuous phase transition, the long range behavior of the system is independent of the details of the microscopic interactions. For example the magnetic order parameter of the Ising model does not depend on the lattice geometry, although the critical temperature is different for a square, triangular, or an hexagonal lattice. 
When different models share the same set of critical exponents it is said that they belong to the same universality class.

In quantum gravity, the conceptual difficulty in extending the notion of universality is the requirement of ``background independence" because it implies that the geometrical structure of the spacetime cannot  play any role in the definition of the microscopic degrees of freedom.

The strategies proposed so far to quantize gravity have different ways of dealing with this issue \cite{kiefer,A,R,T}.
For example, in string theory, although the background independence is not  manifest  at a perturbative level, it should be realized non-perturbatively via AdS/CFT. On the other hand, in loop quantum gravity, this requirement  is satisfied from the very beginning, at least at a formal level, and in the  asymptotic safety program \cite{wein,mr,percadou,oliver1,frank1,oliver2,oliver3,oliver4,souma,frank2, prop,oliverbook,perper1,codello,litimgrav,frankmach,BMS,oliverfrac,jan1,jan2,max,livrev,nagi} background independence is dynamically achieved via the background field method \cite{abbott}.
From this point of view, conformally reduced gravity, a scalar analogous of gravity in which the only propagating degree of freedom of the metric is the conformal factor, is an important theoretical laboratory to understand  the issue of background independence and renormalizability in this context.
In fact, although in classical general relativity the conformal factor is not a propagating degree of freedom, in quantum gravity its fluctuations dominate the path-integral because of the conformal factor instability 
and can become the most important ones both in the ultraviolet (UV) region, close to the non-gaussian fixed point (NGFP) \cite{wein,mr}, and in the infrared sector (IR), below the gaussian fixed point (GFP).

The idea of considering this privileged point of view in order to better understand the structure of the UV critical manifold of quantum gravity was first put forward in \cite{creh1} and \cite{creh2}. In the latter paper a non-perturbative flow equation in the so-called ``local potential approximation" (LPA) has been derived for the first time and further investigated in \cite{elisa} within the framework of the bimetric truncation.
Moreover, in \cite{crehroberto} the contribution of the trace anomaly and the $R^2$ term have been considered.

Important points which deserve to be better investigated in this approach are the dependence of the critical quantities on the threshold functions and the structure of the 
renormalization flow beyond the simple conformally reduced Einstein-Hilbert (CREH) truncation. In the first case it is interesting to investigate the impact of the reduced degrees of freedom as a function of the cutoff structure.
In particular we would like to discuss  the possibility of determining  an ``optimal cutoff" for which the difference between the calculation of the universal properties performed in the full Einstein-Hilbert model and in the CREH approximation is minimal. 
In the second case it is important to study the evolution of admissible initial data that are not necessarily of the CREH form in order to see if non-local contributions to the renormalized Lagrangian significantly deform the structure of the UV critical manifold \cite{frank2}.

The role of non polynomial truncations can be significant also in the deep infrared region.
For instance in the more familiar $\lambda\, \phi^4$-theory in the broken phase, a first-order phase transition occurs in three dimensions, where the inverse susceptibility is not continuous in the $k\to 0^+$ limit. In this case the finite jump of the renormalized mass as a function of the field strength is not accessible to the standard $\bet$ function approach; moreover only with a robust and accurate numerical integration scheme of the flow equation is it possible to recover the convexity property of the free energy in the thermodynamic limit \cite{bonlac,hrt}.

In this work, in particular,  we shall use the proper-time flow equation which has been extremely successful in the calculation of the critical exponents  in the 3-dimensional Ising model \cite{propref} and in quantum gravity \cite{prop}. The main reason to employ this flow equation is that it is rather simple to implement different threshold functions and to interpolate between a sharp momentum cutoff and a sharp proper-time cutoff, as we shall see.

A question that naturally arises in the context of conformally reduced gravity is if a phase of non-zero mean conformal factor $\langle \chi \rangle \not = 0$ takes place at low energy: this symmetry breaking phenomenon could be interpreted as a phase of broken diffeomorphism invariance where $\langle g_{\mu\nu} \rangle \not = 0$ and the spacetime geometry naturally emerges as a low-energy phase.
We shall show that, although the IR flow of the LPA cannot be properly determined for a continuous set of initial data, it is nevertheless possible to study this symmetry breaking phenomenon as an inverse problem, within the LPA approximation. In particular, we shall find a new class of UV fixed potentials which evolves towards a low-energy phase where the diffeomorphism invariance is spontaneously broken.

The structure of the paper is the following: a derivation of the proper-time flow functional equation is discussed in section II by means of a background ``independent" blocking procedure. The fixed points and critical exponents are computed in section III for various classes of threshold functions and for different projection in the background metric.  The results are then compared with those obtained in the full Einstein-Hilbert (EH) model. Section IV includes a new numerical discretization scheme for the flow equation in the LPA  and describes the possibility of having a phase of broken diffeomorphism invariance at low energy. Section V is devoted to the conclusions and in the Appendices the numerical results and explicit expressions of the $\bet$ functions for various regularization schemes and spacetime dimensions are presented.

%----------------------------------------------------------------------------------
%---   SECTION II: Wilsonian action for the conformal factor   ---
%----------------------------------------------------------------------------------
\section{Wilsonian action for the conformal factor}
%----------------------------------------------------------------------------------
In this section we shall introduce the concept of the Wilsonian action for the conformal factor by means of a constraint average field; this derivation is in fact closer to the statistical mechanical point of view than the standard approach based on the Schwinger functionals. It has the advantage of dealing with a quantity that can be directly computed in  Monte Carlo simulations since the introduction of an external current is not necessary \cite{ora} in this case.

Let $\SC[\chi]$ be the action for the fundamental field $\chi(x)$ that we write as $\chi(x) = \cb(x) + f(x)$ where $\cb(x)$ is a non-dynamical background field and $f(x)$ the dynamical (fluctuating)  field, and let $\gat$ be a rigid  reference metric defined on a Euclidean manifold in $d$ dimensions.
The Wilsonian action in the presence of the background field can be formally defined as
\be\label{blocked}
e^{-S_k[\bati;\, \cb  ]} = \int D[f] \, \delta \big (f_k -\bati \big )\, e^{-\SC[ \cb+f]} \, ,
\ee
where $f_k(x)$ is an averaged fluctuation field given as 
\be
f_k(x) = \int d^dy \; \sqrt{\hat g} \, f(y) \; \rho_k(y, x; \cb)   \, ,
\ee
and where  $\rho_k(x,y; \cb)$ is a smearing kernel with the properties  of being
\begin{itemize}
\item \hspace{-0.1em}symmetric: $\rho_k(x,y; \cb)\equiv \rho_k(y,x; \cb)\, ,$ 
\item \hspace{-0.1em}normalized: $\int d^dy \; \sqrt{\hat g} \, \rho_k(y, x; \cb) = 1\, ,$
\item \hspace{-0.1em}idempotent:  $\int d^dy \; \sqrt{\hat g} \, f_k(y) \; \rho_k(y, x; \cb) = f_k(x)\, .$
\end{itemize}
The last relation simply implies that the average of an average field is again an average field \cite{wette91} (see \cite{boneu} for a general discussion on smearing kernels in Riemannian spaces).
Its explicit expression in terms of $\cb$ dependence does not need to be specified at this level.

In this formalism $\chi$ plays the same role of the microscopic metric $\gamma_{\mu\nu}$ in the full theory. 
In the complete framework a background metric $\bat$ is chosen in order to perform the actual calculations, and the fluctuations $h_{\mu\nu}$ are quantized non-perturbatively around this background which will be dynamically determined by the requirement that the expectation value of the fluctuation field vanishes, $\langle h_{\mu\nu}\rangle\equiv \bar h_{\mu\nu}=0$.  
Any physical length must then be proper with respect to the background metric $\bat$. 
In the conformally reduced theory the expectation values $\bar f\equiv \langle f\rangle$ and $\phi \equiv \langle \chi \rangle=\cb +\bar f$ are the analogs of $\bar h_{\mu\nu}\equiv \langle h_{\mu\nu} \rangle$ and $g_{\mu\nu} = \langle \gamma_{\mu\nu} \rangle = \bar g_{\mu\nu}+\bar h_{\mu\nu}$ in the full theory.

The central idea of the conformal field quantization is to employ the background metric 
\be
\bat \equiv \cb^{2\nu}\, \gat
\ee
in constructing the smearing function $\rho_{\kb}(x,y) \equiv \rho_{k}(x,y; \chi_B)$ via the spectrum of $-\babox$, being $\kb$ and $k \equiv \ka$, respectively the momentum operators built with the background metric $\bat$ and the fixed metric $\gat$, $\nu \equiv {2}/({d-2})$ and $d$ is the space-time dimension.
The reference metric $\gat$ plays no dynamical role in this process but it is fixed to perform the actual calculation, while all the dynamical fields are spectrally decomposed using the basis of the $-\babox$ eigenfunctions whose eigenvalues satisfy 
\be
\label{kk}
\kb^2= \cb^{-2\, \nu}\, k^2
\ee
in the case of a  constant $\cb$.
The $\delta$-function kernel in (\ref{blocked}) 
\be
\delta \big (f_k -\bati \big ) =\prod_x \delta(f_k(x) -\bati (x)) 
\ee
reduces to $\delta(f-\bati)$ in the $k\rightarrow \infty$ limit and projects on the zero-momentum mode in the  $k \to 0^+$ limit instead.

Therefore at $k=0$ the blocked action (\ref{blocked}) coincides with the effective potential, namely, the non-derivative part of the effective action, but for $k\not = 0$ the functional described in (\ref{blocked}) is an effective action for the ``low-energy'' modes with momentum $p<k$.
The relation with the standard effective average action for the conformal factor defined in \cite{creh1,creh2} can be obtained by noticing that the expectation value of the blocking field $\bati(x)$, which is defined as
\begin{widetext}
\ba\label{blocked2}
\langle\, \bati\, \rangle = && \int D[\bati(x)] \, e^{-S_k[\bati(x);\, \cb (x) ]} \bati(x) = \int D[\bati (x)]\, D[f(x) ]\,   \delta \big (f_k(x) - \bati (x) \big ) e^{- \SC[\cb(x) + f(x)] } \, \bati(x)  = \nonumber \\
= && \int D[f(x)]\, e^{- \SC[\cb(x) + f(x)] } f_k(x) = \langle \, f_k(x)\, \rangle \, ,
\ea
\end{widetext}
can be rewritten, after having introduced the source $J(x)$, as
\be
\langle\, \bati\, \rangle = \frac{1}{\sqrt{\hat g}}\frac{\partial W_k[J; \cb]}{\partial J(x)}\Big |_{J = 0} \, ,
\ee
where 
\be\label{greenf}
W_k[J; \cb] = \int d^dx\, \{J(x)\, \bati(x) - S_k[\bati; \cb ]\} \, . 
\ee
The (\ref{greenf}) corresponds to the standard generating functional $W_k[J; \chi_B]$ defined in \cite{creh1}. 

The $\delta$-function constraint in (\ref{blocked}) can be conveniently evaluated in the momentum space and the ``background-blocked'' action $S_{\kb} \equiv S_k[\bati; \cb]$ can be explicitly computed in the one-loop approximation.
The difference $\Delta S_{\kb} = S_{\kb+\delta \kb} - S_{\kb}$ can then be evaluated in the infinitesimal momentum shell between $\kb$ and $\kb+\delta \kb$, where $\kb$ is the ``proper" momentum operator built with the background metric $\bat$. A functional flow equation is finally obtained by taking the $\delta \kb \rightarrow 0$ limit and performing a renormalization group improvement of the resulting expression \cite{prop}. 
After this step is accomplished, the ``background-independent" flow is obtained expressing all the running ``proper" momenta in terms of the reference energy scale $k$.
Rewriting the (regularized) one-loop contribution in the Schwinger ``proper-time" formalism one finds
\be\label{+}
\partial_t\, {S}_k [\bati; \chi_B] =  - \frac{1}{2}\,  {\rm Tr} \int_0^{\infty} \frac{ds}{s}
\, \partial_t\, \tau_{k}\, \exp \Big\{ -s \frac{\delta^2 S_k [\bati; \chi_B]}{\delta {\bati}^2} \Big\} \, ,
\ee
where $t \equiv \log(k)$ is the RG time and $\tau_k\equiv \tau_k[\chi_B]$.
The important difference between this type of functional ``proper-time" flow equation and the version used in earlier investigations \cite{propref} is that the trace in (\ref{+}) is here computed by means of  the representation provided by the spectrum of  $-\babox$, 
\be
\overline{{\rm Tr}} [A] \equiv \int d^dx\, \sqrt{\bar{g}}\,\, \langle x | A | x \rangle = \int d^dx\, \sqrt{\hat{g}}\,\cb^{d \nu}\, \langle x | A | x \rangle\, .
\ee
The precise relation between the ``proper-time" flow equation and the exact flow equation \cite{avact} has been extensively clarified in \cite{ergprop}.
For actual calculations  we shall use the one-parameter family of smooth cutoffs $\tau_k\equiv \tau_k^{n}$ that has been widely used in the literature \cite{propref}, whose explicit expression reads
\be\label{3}
\tau_k^n(s) = \frac{\Gamma(n, s\, {\cal Z}\, n\,k^2\, \cb^{2 \nu} )  - \Gamma(n, s\, {\cal Z}\, n\, \Lc^2\, \cb^{2 \nu}) }{\Gamma(n)} \, .
\ee
Here $n$ is an arbitrary real, positive parameter that controls the shape of the $\tau_k^n$ in the interpolating regions, and $\Gamma(\alpha,x)=\int^\infty_x dt \; t^{\alpha-1} e^{-t}$ denotes the incomplete Gamma-function. Furthermore, ${\cal Z}$ is a constant which has to be adjusted to make sure that the eigenvalues of $-\babox$ are cut off around $\sim k^2$  rather than $\sim k^2/{\cal Z}_a$ \cite{prop}.
Therefore derivative $\partial_t\, \tau_k^n(s)$ in (\ref{+}) explicitly reads
\ba\label{4}
&&\partial_t \tau_k^n(s) \equiv \lim_{\delta k\rightarrow 0}  \, k \, 
\frac{\tau^n_{k+\delta k}(s) - \tau^n_k(s)}{\delta k}=\nonumber\\[2mm]
&& -\frac{2}{n!}\; ({\cal Z}\, s\, k^2\, n\, \cb^{2 \nu})^{n} \exp (-{\cal Z}\, s\, k^2\, n\, \cb^{2 \nu}) \, ,
\ea
with $n>d/2$. 
For $n=d/2$ the kernel (\ref{4}) does not regulate completely the UV because the proper-time integral requires a field independent (vacuum) contribution to be subtracted from the right-hand side of Eq.(\ref{+}).
On the other hand for $n>d/2$ this class of regulators allow us to take the formal limit $n\rightarrow \infty$ which coincides with a sharp-cutoff introduced in the proper-time cutoff \cite{Flore}. To conclude this section we would like to remark that for $n=d/2+1$ our regulator corresponds to the ``optimized cutoff" introduced in \cite{optim}.

%----------------------------------------------------------------------------------
%---               SECTION III: Polynomial truncations                     ---
%----------------------------------------------------------------------------------
\section{Polynomial truncations}  
%----------------------------------------------------------------------------------
In this section we shall discuss  the structure of the NGFP obtained by the flow equation (\ref{+}) as a function of the cutoff parameter $n$ for different reference topologies.  

It is important to remark that, at variance with the well-known definition of the path-integral for quantum-gravity based on the sum over all possible metric/topologies, in our case the use of different topologies is only a technical device to project an infinite-dimensional functional flow equation in a finite dimensional  theory space where only the flow of $\sqrt{g} R$ and $\sqrt{g}$ operators is considered. From this point of view our approach has nothing to do with a calculation performed in the Gibbons-Hawking spirit. Neither are we expanding the graviton propagator in inverse powers of momentum/curvature. 
On the contrary the (unprojected) functional flow equation is, by construction, independent on the topology (see \cite{mr}) and the same property is shared by the flow equation for the conformal factor (see e.g. \cite{creh1,creh2}).
However because the irrelevant operators of the NGFP have a  different impact on the renormalized flow at the zeroth order of the gradient expansion (spherical projection) and at first order (flat projection), the universal quantities will show this residual scheme dependence.

Let us then assume that the field dependence of the blocked action is completely encoded in a relation of the type $\phi=\cb +\bati$, where $\phi$ is a blocked field, so that $S_k$ is a local function of $\phi$. 
An important example of this approximation is the conformally reduced Einstein-Hilbert truncation. 
It is obtained by setting $g_{\mu \nu} =\phi^{2 \nu} \, \,\widehat{g}_{\mu \nu}$ in the Euclidean Einstein--Hilbert action \cite{jackiw}
\be \label{2.1}
S_k^{\text EH}[g_{\mu \nu}] = - \frac{1}{16 \pi} \,\int \!\!\mathrm{d}^d x~\sqrt{g\,} \, G_k^{-1}\bigl( R (g) - 2 \, \Lambda_k \bigr)
\ee
so that the standard formulas for Weyl rescalings yield 
\ba\label{2.4}
S_k[\phi]=
\int \!\! \mathrm{d}^d x\, \sqrt{\widehat{g} \,} \,\,
Z_k \, &\Big(& \tfrac{1}{2} \, \widehat{g}\,^{\mu \nu}  \partial_\mu \phi \, 
\partial_\nu \phi
+ \tfrac{1}{2}\,  A (d)\,  \widehat{R} \, \phi^2 + \nonumber\\[2mm]
&&  - 2\, A(d) \, \Lambda_k \, \phi^{\frac{2d}{(d-2)}} \Big) \, ,
\ea
where ${\widehat R}\equiv {R(\widehat g)}$ and
\be
Z_k = -\frac{1}{2\,\pi\, G_k}\frac{d-1}{d-2}, \;\;\;\;\;\; A(d) = \frac{d-2}{8\,(d-1)}.
\ee
In particular, the second functional derivative of (\ref{2.4}) reads 
\ba
\label{second}
S^{(2)}_k [\phi] =&& Z_k\, \Big(- \gabox + 2\, A(d)\, \widehat{R} +\\[2mm]
&& - 2\, A(d)\, B(d)\, \Lambda_k\, \phi^{\frac{2d}{(d-2)} - 2} \Big)\, \delta^d(x,y) \, ,\nonumber
\ea
where $\delta^d(x,y)$ is the Dirac delta function in $d$ dimensions in the fixed metric space endowed with the $\gat$ metric, and
\be
B(d) = \frac{2d}{d-2}\left(\frac{2d}{d-2} -1\right)\, .
\ee

%------------------------------------
%             S^d topology
%------------------------------------
\subsection{$S^d$ topology}
%------------------------------------
Let us first consider the topology of the $d$-dimensional sphere $S^d$. 
In this case the curvature of the reference metric $\gat$ is constant and the running of the dimensionless coupling $g_k = G_k\, k^{d-2}$ can be obtained from the $\phi^2$ term setting $\cb(x) = constant$ in the action (\ref{2.4}), so that
\be
S_k^{S^d}[\cb] = 
\int \!\! \mathrm{d}^d x\, \sqrt{\widehat{g} \,} \,
Z_k \, \Big(\tfrac{1}{2}\,  A (d)\,  \widehat{R} \, \cb^2  - 2\, A(d) \, \Lambda_k \, \cb^{\frac{2d}{(d-2)}} \Big)\, .\vspace{1em}
\ee
The trace on the background metric can be computed using the Seeley-Gilkey-deWitt heat kernel expansion up to linear terms in the reference curvature by using the Mellin transform $Q_n$, so that
\ba\label{heat}
&&\overline{{\rm Tr}}\, [\, W(-\gabox)] = \\[2mm]
&&\frac{\cb(x)^{d\,\nu}}{(4\,\pi)^{d/2}}\, \left \{ Q_{\frac{d}{2}} \, \int d^d x\, \sqrt{{\hat g}} + \frac{1}{6}\,Q_{\frac{d}{2}-1}\, \int d^d x\, \sqrt{{\hat g}}\, {\hat R} \right \}   \, , \nonumber
\ea
with
\be
Q_n\, [\, W(-\gabox)] = \frac{1}{\Gamma[n]} \int^\infty_0 \, q^{n-1} W(q)\, dq \, . 
\ee
In particular, using the relation $- \gabox = - \babox\, \cb^{ 2\, \nu}$, $W(-\gabox) = e^{- s\, Z_k\,(-\gabox))} = e^{- s\, Z_k\, \cb^{2\, \nu} (-\babox))}$. Therefore
\begin{subequations}\label{zz}
\ba 
&&\hspace{-1.2em}Q_{\frac{d}{2}}= \frac{1}{\Gamma\left [\frac{d}{2} \right ]} \int dq \, q^{\frac{d}{2}-1} \, e^{-s Z_k \cb^{2\, \nu} q} = \frac{1}{(s \,Z_k \,\cb^{2\, \nu})^{\frac{d}{2}}} \, , \\[2mm]
&&\hspace{-1.2em}Q_{\frac{d}{2}-1}= \frac{1}{\Gamma\left [\frac{d}{2} -1\right ]} \int dq \, q^{\frac{d}{2}-2} \, e^{-s Z_k \cb^{2\, \nu} q} = \frac{1}{(s \,Z_k \,\cb^{2\, \nu})^{\frac{d}{2}-1}}\, . \nonumber\\[2mm]
\ea
\end{subequations}
By inserting expression (\ref{second}) and (\ref{4}) in the flow equation (\ref{+}) and using (\ref{heat}) with (\ref{zz}), the coefficients of the $\cb^2$ and $\cb^{2d/(d-2)}$ terms due to the renormalized flow 
of the $\sqrt{\gat} \, {R\, (\gat)}$  and $\sqrt{\gat}$ operators are easily identified.
At last the $\bet$ functions for the dimensionless running Newton constant $g_k$ and the dimensionless Cosmological constant $\lambda_k = \Lambda_k\, k^2$ can be obtained with the introduction of  the ``anomalous dimension'' $\eta \equiv k\, \partial_k\, \ln G_k$, so that 
\vspace{-0.5em}
\be
\beta_g(g,\lambda) \equiv k\, \partial_k\, g_k = (d - 2 + \eta)\,g_k
\ee
and $\eta \equiv \eta(g_k, \lambda_k)$ in general. In four dimensions we have 
\begin{subequations}\label{beta1}
\ba
\label{b1}
\beta_g = &&g_k\, \left( 2 - \frac{g_k}{(n - 2\, \lambda_k )^{n-1}}\frac{n^n\, \Gamma[n-1]}{6\, \pi\, \Gamma[n] } \right) \, , \\[2mm]
\beta_\lambda = &&\lambda_k \left( - 2  - \frac{g_k}{(n - 2\, \lambda_k )^{n-1}}\frac{n^n\, \Gamma[n-1]}{6\, \pi\, \Gamma[n] } \right) +\nonumber\\[2mm]
&& + \frac{g_k}{(n- 2\, \lambda_k)^{n-2}}\frac{n^n\, \Gamma[n-2]}{2\, \pi\, \Gamma[n]}\, . 
\label{b2}
\ea
\end{subequations}
The expressions for the $d$-dimensional $\bet$ functions are listed in the Appendix B, together with the limiting cases $n \to d/2$ and $n \to \infty$.

%-------------------------------------------------
%                   R^d topology
%-------------------------------------------------
\subsection{$\mathbb{R}^d$ topology}
%-------------------------------------------------
In the case of a flat $\mathbb{R}^d$ topology the scalar curvature  of the reference metric vanishes, constraining the \textit{quadratic term} of the action (\ref{2.4}) to be zero.
In order to extract the beta function from the flow equation (\ref{+}) it is convenient to consider a general truncation of the type\\
\vspace{-1em}
\be\label{r4}
S_k[\phi] = \int \!\! \mathrm{d}^d x~ \sqrt{\widehat{g} \,} \, \left(\tfrac{1}{2} \, \widehat{g}\,^{\mu \nu}\,   Z_k\, \partial_\mu \,\phi \, \partial_\nu\, \phi\, +V_k(\phi) \right) \, ,\\
\ee
where $V_k(\phi)\, =\, Z_k\, U_k(\phi)$  and employ a derivative expansion around an homogeneous background plus a fluctuation, so that $\phi=\cb + \bati(x)$. In this case we have 
%Therefore
\ba
&&\hspace{-0.8em}S_k[\phi]=  \int d^dx \sqrt{\widehat{g}} \, \Big \{- \frac{1}{2}\, Z_k\, \bati(x)\, \widehat{\square}\, \bati(x) + V_k[\chi_B] +\\[2mm]
&&\hspace{-0.8em}+ V_k'[\chi_B]\, \bati(x) + \frac{1}{2}\, V_k''[\chi_B]\, \bati(x)^2 + \mathcal{O}(\bati(x)^3) + \mathcal{O}(\partial^4\, \bati) \Big \}\nonumber \, .
\ea
Therefore,
\ba\label{bef}
&&\partial_t \, S_k[\bati(x)] =\\
&& -\frac{1}{2}\, \int d^dx\, \sqrt{\hat{g}}\,\, \chi_B^{d\,\nu}\,  \int \frac{ds}{s}\,\partial_t \tau_k^n\, \langle\, x\, |\, e^{- s\, (K+\delta K)} \,|\,x\, \rangle\, , \nonumber
\ea
where
\begin{subequations} 
\ba\label{KK}
&&K= - Z_k\, \widehat{\square} + V_k''[\chi_B] \, ,\\[2mm]
&& \delta K =  V_k'''[\chi_B] \bati(x) + \frac{1}{2}\, V_k''''[\chi_B]\, \bati(x)^2 \, .
\ea
\end{subequations}
The trace in (\ref{bef}) can be 
evaluated in a background-independent way by means of an integration in momentum space over the eigenvalues $\bar{p}^2$ of the Laplacian built from the background metric $\bat$, inserting in (\ref{bef}) the identity $\int d^d \bar{p}\, |\,\bar{p}\,\rangle \langle\, \bar p\, | = \mathbb{I}\, (2\, \pi)^d$ and using in (\ref{KK}) the substitution $- \gabox \to - \babox\, \cb^{ 2\, \nu}$.
In order to disentangle the trace in (\ref{bef}) a Baker-Campbell-Hausdorff expansion of the heat kernel is performed, so that
\begin{widetext}
\be
\partial_t \, S_k[\bati(x)] =-\frac{1}{2}\, \int d^dx\, \sqrt{\hat{g}}\,\,\cb^{d\,\nu}\, 
\int \frac{d^d\bar{p}}{(2\, \pi)^d} \int\, \frac{ds}{s}\, \partial_t \tau_k^n(s)\, 
\langle\, x\, |\, \bar p\, \rangle \langle\, \bar p\, |\, e^{- s\, K} (1 - s\, \delta K + \frac{s^2}{2!} \{ [ \delta K, K] +  \delta K^2\} + \dots\,) |\, x\, \rangle \, ,
\label{fr4}
\ee
\vspace{0.2em}
\end{widetext}
where the dots stand for the higher order terms in the $s$ expansion of the exponential and 
\be
\langle\, x\, |\, \bar p\, \rangle = e^{- i \, \bar p \, x}\, .
\ee
The matrix elements of the expanded heat kernel can then be calculated ordering the operators by means of the commutation rule   
\be
[\bar p_\mu, \bati(x)] = - i \, \partial_\mu\, \bati(x)\, .
\ee
It is then straightforward to identify the coefficients of the $V_k$ and $Z_k$ terms, obtaining the following set of coupled equations:
\begin{subequations}\label{beh}
\ba\label{beh1}
&&k\, \partial_k\, V_k = M\, (k^2\, \cb^{2\, \nu})^{\frac{d}{2}}  \frac{1}{\left (1+ \frac{V_k''(\cb)}{k^2\, n\, Z_k\, \cb^2} \right)^{n-\frac{d}{2}}} \, ,\\
&&k\, \partial_k\, Z_k = N\, (k^2 \cb^{2 \, \nu})^{\frac{d}{2}-3} \frac{\left (V_k'''/Z_k\right )^{2}}{ \left ( 1+ \frac{V_k''(\cb)}{k^2 nZ_k \cb^2} \right)^{n+3-\frac{d}{2}}}\, , \hspace{3em}
\label{beh2}
\ea
\end{subequations}
where 
\begin{subequations}\label{beh3}
\ba
&&M = \left ( \frac{n}{4\pi} \right)^\dm\frac{\Gamma(n-\dm)}{\Gamma(n)} \, ,\\
&&N =  \frac{(d-2\, (n+1)) (d-2\, (n+2))}{24\,  d \, n^2} \, \left ( \frac{n}{4\pi} \right)^\dm\frac{\Gamma(n-\dm)}{\Gamma(n)} \, .\nonumber\\[2mm]
\ea
\end{subequations}
The $\bet$ functions for the dimensionless couplings of the CREH truncation are then obtained introducing a polynomial ansatz for the dimensionful potential $U_k$ of the type
\be 
U_k[\cb] = -\, k^2 \, \frac{\lambda_k}{6} \; \cb^4 \, ,
\ee
so that one obtains in four dimensions the coupled set of equations
\begin{subequations}
\label{beta2}
\ba
\beta_g = &&g_k\, \left(2 - 2\, \frac{g_k\, \lambda_k^2}{(n - 2\, \lambda_k )^{n - 1}} \frac{\Gamma[n+1]\, n^n}{9\, \pi\, \Gamma[n]} \right) \, , \,\,
\ea
\ba
\beta_\lambda = && \frac{g_k}{(n - 2\, \lambda_k )^{n-2}}     \frac{ \Gamma (n-2)\, n^n}{2\, \pi\,  \Gamma[n]} +\\[2mm] 
&&+ \lambda_k  \left(- 2 - 2\, \frac{g_k\, \lambda_k^2}{(n - 2\, \lambda_k )^{n - 1}} \frac{\Gamma[n+1]\, n^n}{9\, \pi\, \Gamma[n]} \right)\, . \nonumber
\ea
\end{subequations}
%

%----------------------------------------------------------
%         Fixed points ad linearized flow
%----------------------------------------------------------
\subsection{Fixed points and linearized flow}
%----------------------------------------------------------
The $\bet$ functions (\ref{beta1}) and (\ref{beta2}) vanish both at the GFP located at $\las = \gas = 0$, and at a NGFP defined at $\las\not =0$, $\gas\not = 0$.
The properties of the linearized flow around the NGFP are determined by the stability matrix $B$
\be
B_{\,i j} = \left( \partial_{g_i} \, \bet_{g_j}\right) \Big |_{\{g_{i}\} = \{g_{i}^*\}} \, ,
\ee
$\{g_i\} \in \{g,\lambda\}$, whose eigenvalues $\theta_{1,2}= - \tp\pm i \tpp$ form in general a complex conjugate pair. 
A negative real part of the eigenvalues, i.e. a positive $\tp$ (we will refer to it as the first Lyapunov exponent, following the standard notation used in dynamical systems), implies the stability of the fixed point, while the imaginary part characterizes the spiral shape near the fixed point. 
Our results in four and $d$ dimensions are summarized, respectively, in Table \ref{tab:1} and Table \ref{tab:2} in the Appendix A.

It is clear from Table \ref{tab:1} that also the theory defined by the CREH approximation is asymptotically safe, although the scaling properties are rather different from those obtained from the full EH in \cite{prop}. 
For instance, the critical exponents $\tp$ and $\tpp$ display an $n$-dependence which is stronger in the case of the CREH than for the non-reduced theory, although the quantity $\las\, \gas$ is rather stable in both cases. 

We can quantify the impact of the EH conformal reduction with respect to the full EH theory by defining a $\chi^2$-type of ``distance" in the space of the ``universal'' quantities, by means of
\begin{widetext}
\be
\chi^2(n)= \frac{(\las\gas(C) - \las\gas(E))^2}{ \las\gas(C)^2+\las\gas(E)^2}+\frac{(\tp(C)-\tp(E))^2}{\tp(C)^2+\tp(E)^2} \, ,
+ \frac{(\tpp(C)-\tpp(E))^2}{\tpp(C)^2+\tpp(E)^2}
\ee
\end{widetext}
where ``C" and ``E" stands for CREH and EH, respectively. 

\begin{center}
\begin{figure}
\includegraphics[width=8.5cm]{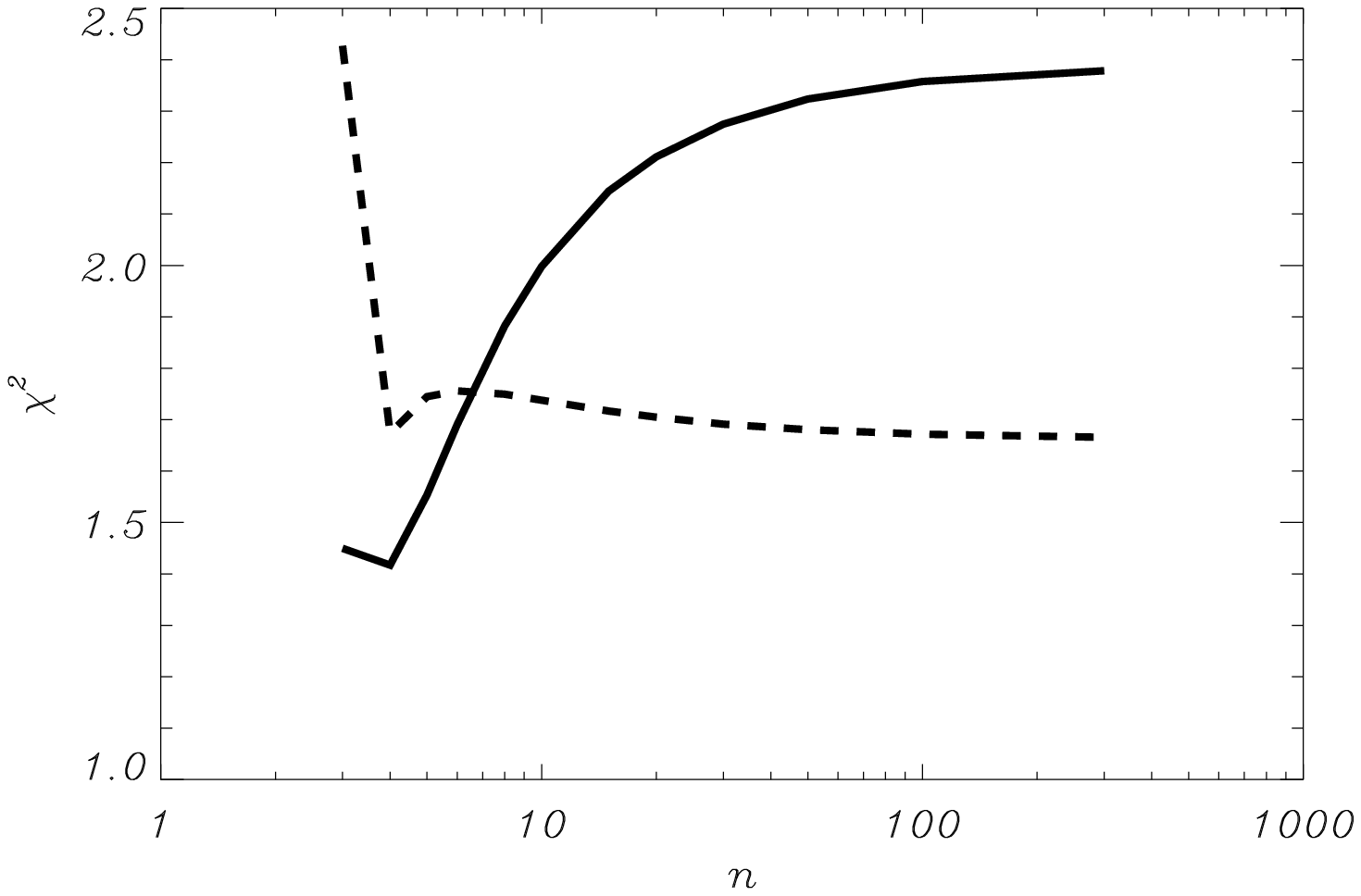}
\includegraphics[width=8.5cm]{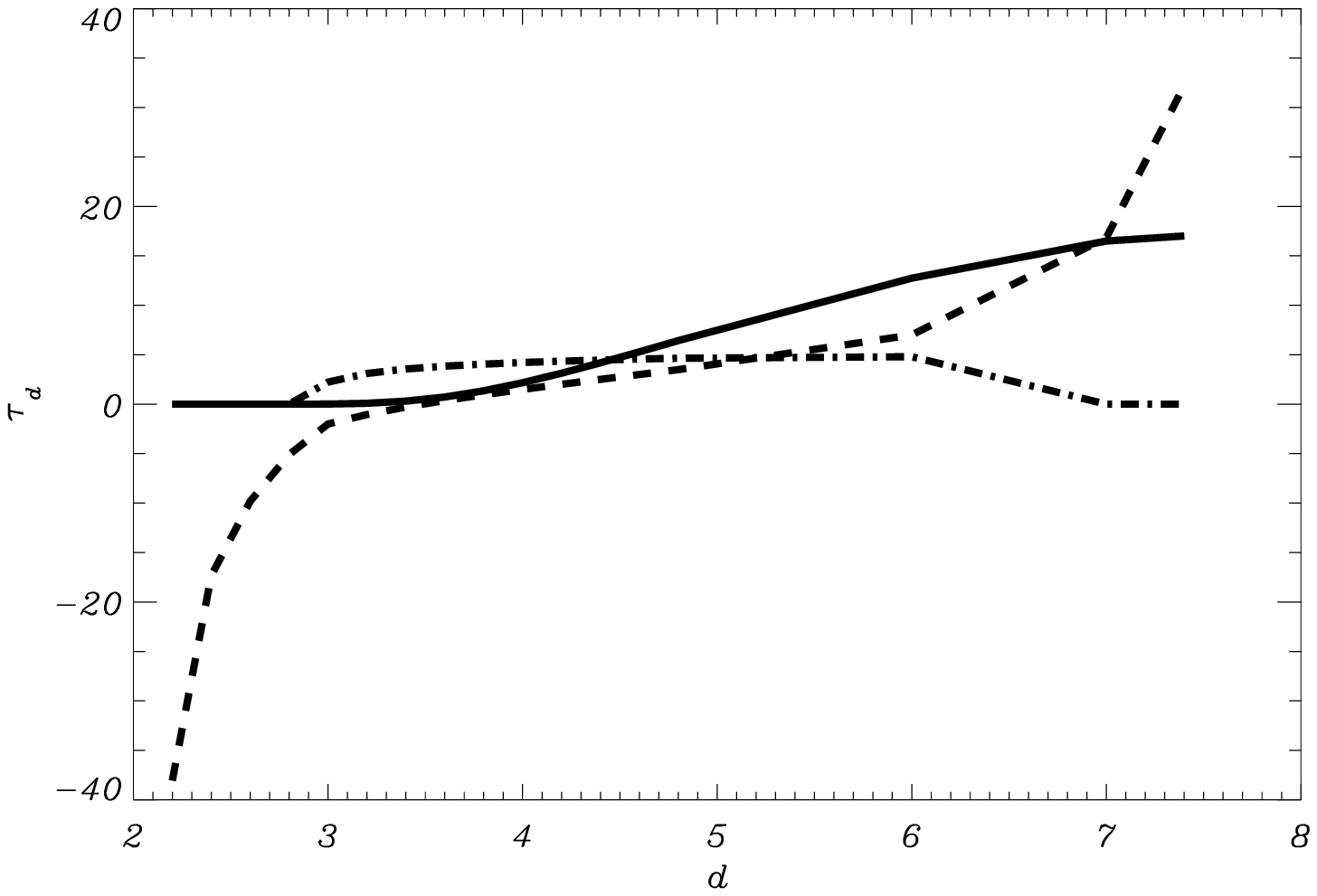}
\includegraphics[width=8.5cm]{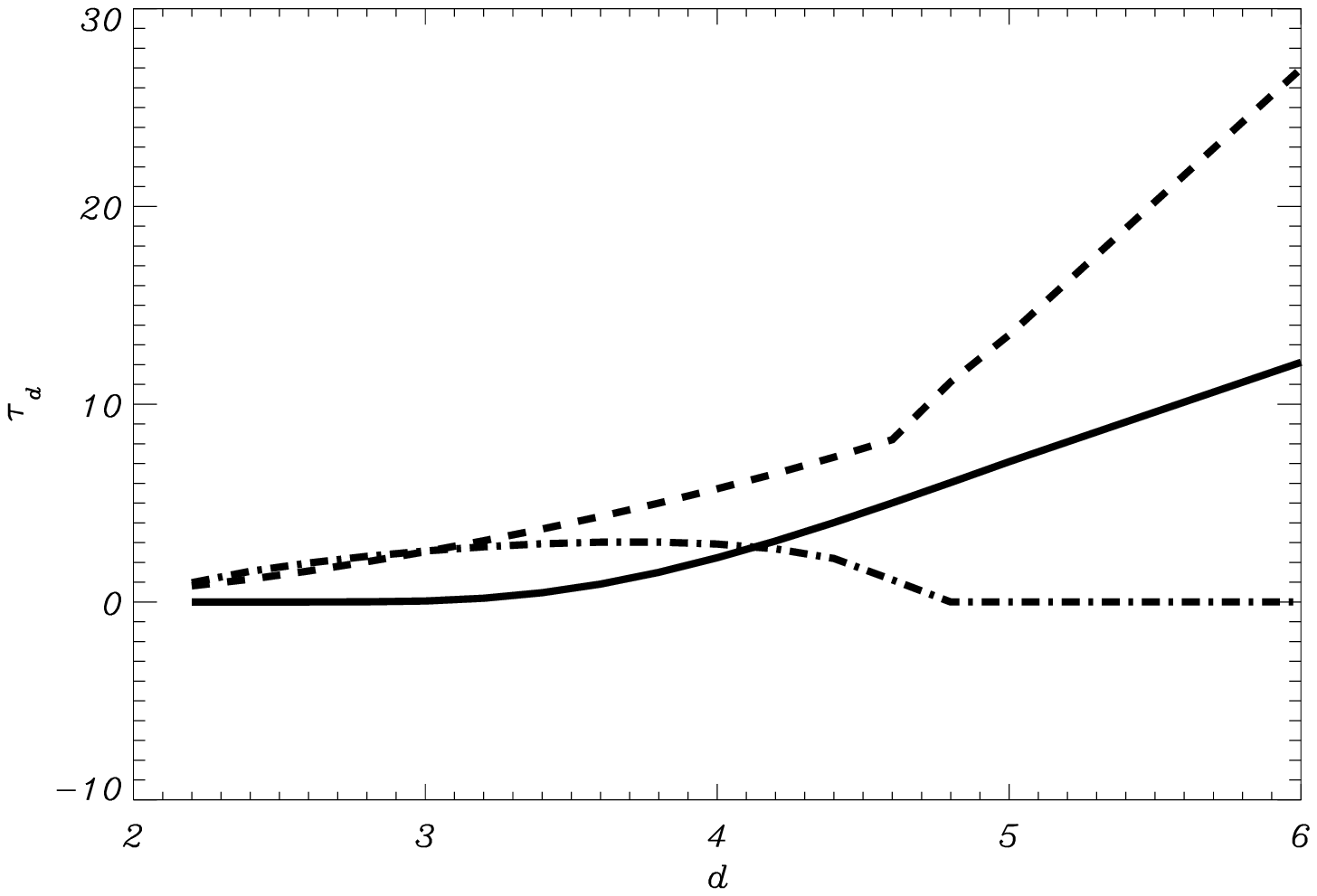}
\caption{
Top: the quantity $\chi^2(n)$ as a function of the cutoff parameter $n$ in the case of $S^4$ projection (solid line) and $\mathbb{R}^4$ (dashed line).
Middle and bottom: the quantity $\tau_d$ (solid line), $\tp$ (dashed line) and $\tpp$ (dotted-dashed line) as a function of the dimension $d$ for $n = 4$ in the case of $S^4$ (middle) and $\mathbb{R}^4$ (bottom) projections.
\label{fig:chi2dim}}
\end{figure}
\end{center}

A plot of this quantity as a function of $n$ is depicted in the upper panel of Fig.(\ref{fig:chi2dim}), for the $S^4$ projection (solid line) and the $\mathbb{R}^4$ projection (dashed line) where it is clear that the minimum is attained for $n=4$ in both cases. On the other hand, in the case of the $\mathbb{R}^d$ projection the scaling properties are much less sensitive to the cutoff parameter $n$, and the $n=\infty$ limit is as good as the $n=4$ case. 

Of particular interest is the $n = \infty$ limit for the $S^d$ topology, in which the first Lyapunov exponent vanishes.
In this case the theory is still UV finite although not asymptotically safe anymore, since now the linearized system is defined by pure imaginary eigenvalues $\pm \tpp$ and every perturbation of the NGFP will evolve in a cyclic trajectory.

\begin{center}
\begin{figure*}
\includegraphics[width = 5.9 cm]{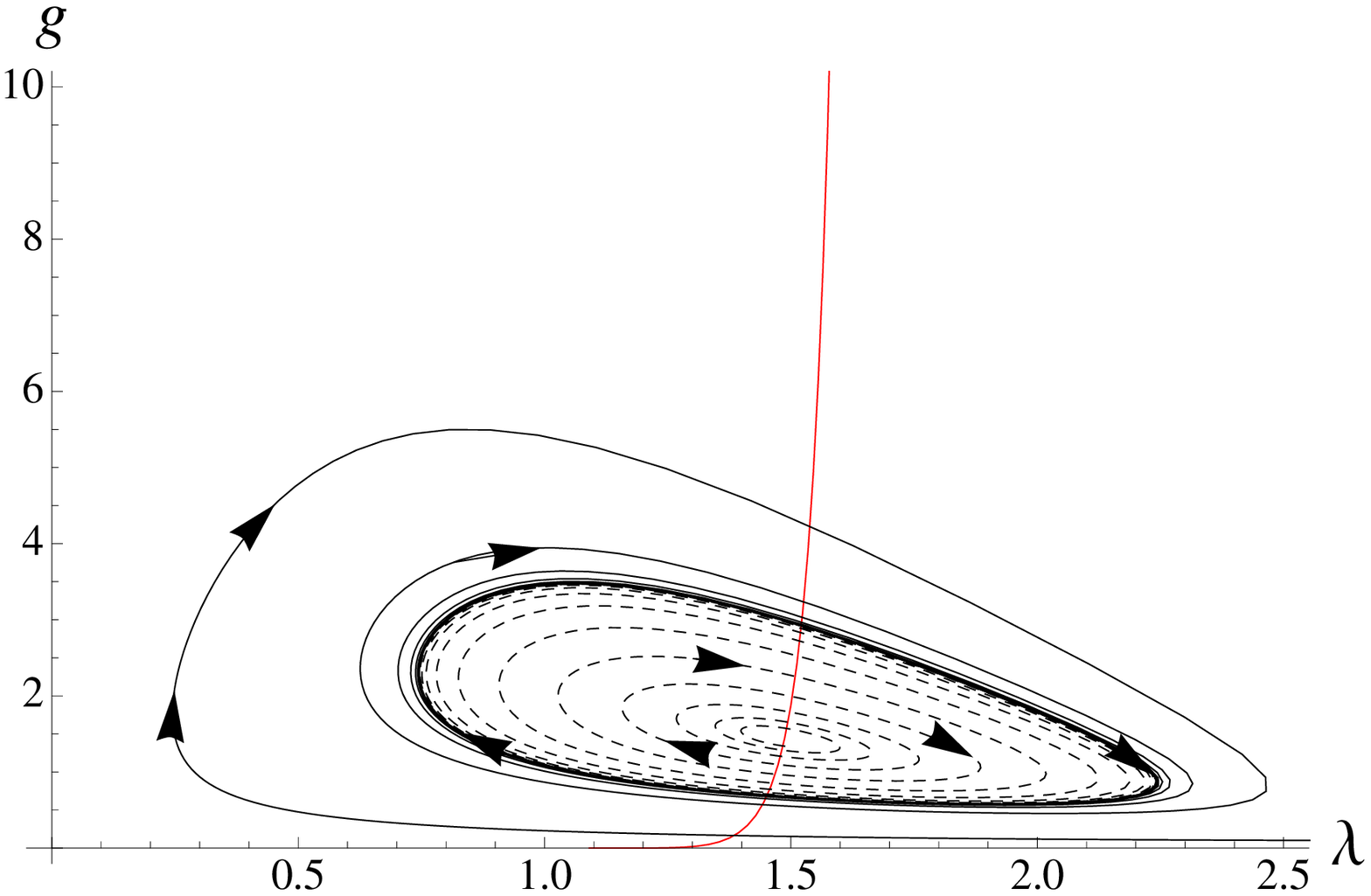}
\includegraphics[width = 5.9 cm]{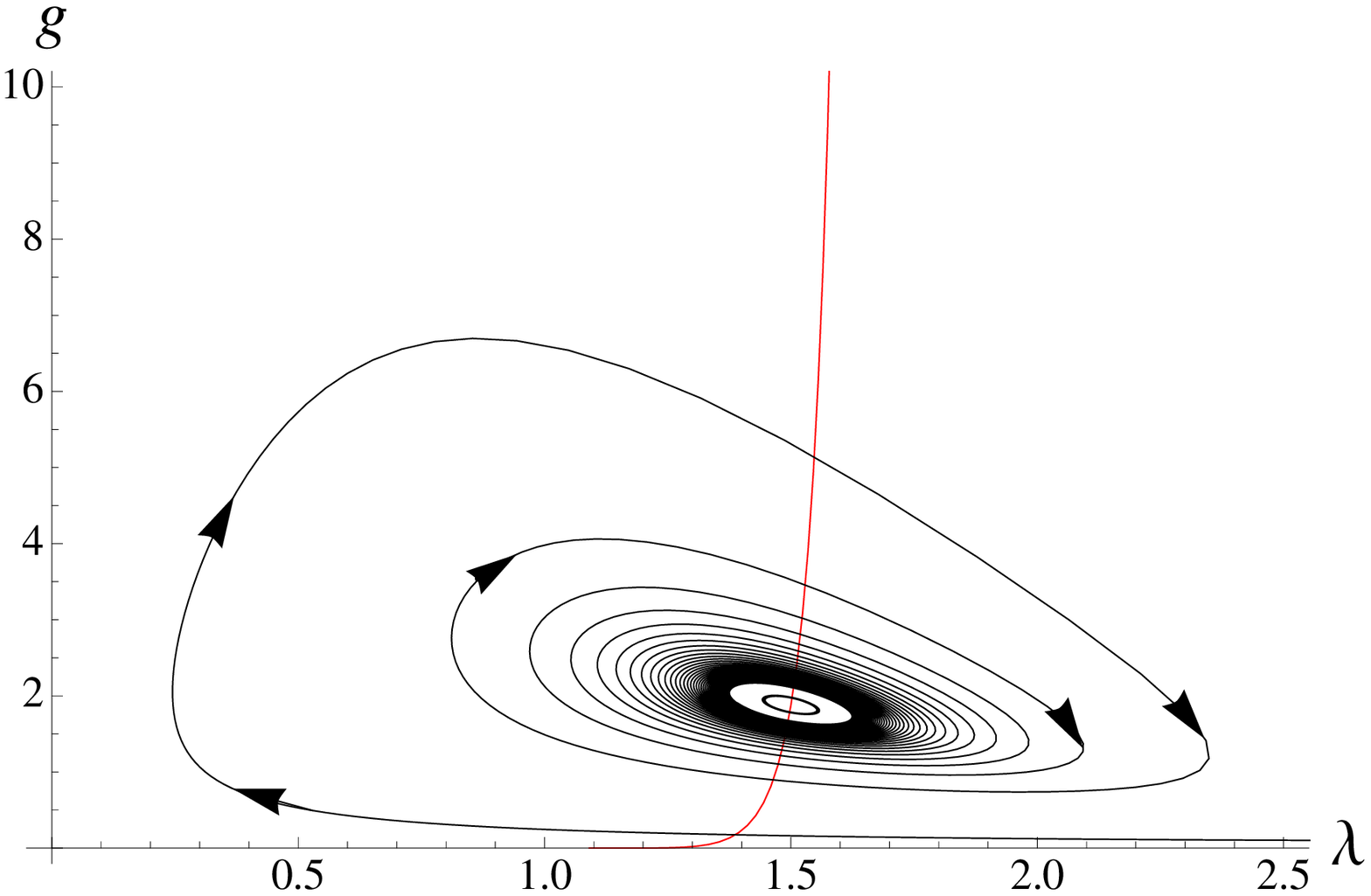}
\includegraphics[width = 5.9 cm]{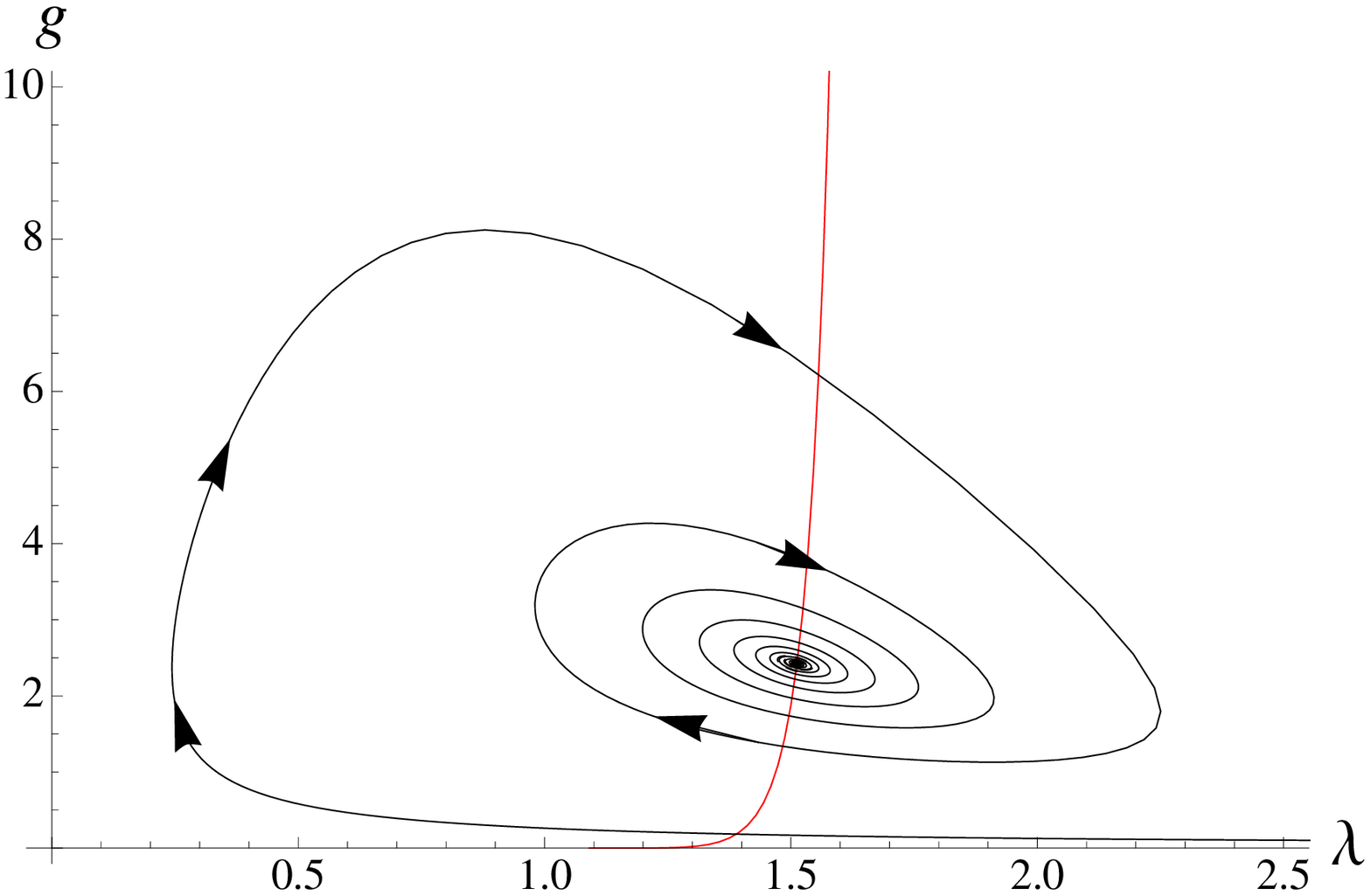}
\caption{Flows in $d$ dimension for the CREH $S^d$ projection. Left: the limit cycle at $d <  d_c$, the dashed line is the repulsive internal flow. Middle: the limit cycle at the critical dimension $d = d_c$. Right: the flow of the UV attractive NGFP at $d >  d_c$. The red line is the location of the FP as a function of the dimension $d$. The plots have been obtained setting $n=\infty$ and, starting from the left, for $d=3.9$, $d=4$, $d=4.1$.}
\label{cycles1}
\end{figure*}
\end{center}

It is also interesting to discuss the scaling properties of the theory in the $S^d$ projection as the dimension is changed. 
This is shown in the middle panel of Fig.(\ref{fig:chi2dim}) for $n=4$ for $\tp$, $\tpp$ and for the dimensionless quantity $\tau_d \equiv \las\, \gas^{2 / (d-2)} = \Lambda_k \, G_k^{2/(d-2)}$. 
The first Lyapunov exponent $\tp$ vanishes for a critical dimension value $d_c$ so that the fixed point undergoes an Hopf bifurcation as the dimension $d$ crosses $d_c$ (represented in Fig.(\ref{cycles1})). 

\begin{figure}
\includegraphics[width = 8 cm]{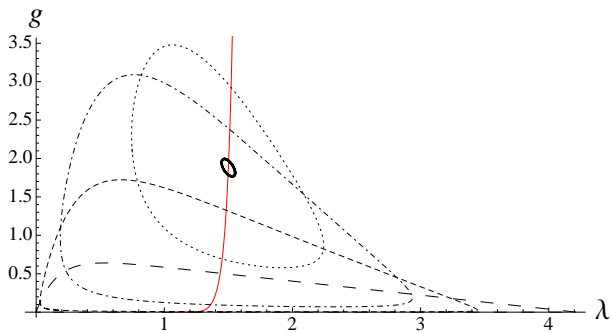}
\caption{The Hopf bifurcation as a function of the dimension $d$ for $n = \infty$ in the $g$-$\lambda$ plane. The solid line is a cycle for an initial value near the FP at $d_c = 4$, the other cycles plotted are at $d = 3.9$ (dotted line), $d = 3.6$ (dotted-dashed line), $d = 3.3$ (dashed line) and $d = 2.9$ (long dashed line). 
The red line is the location of the FP as a function of the dimension $d$.}
\label{cycles2}
\end{figure}

As it is shown in Fig.(\ref{cycles2}), for $d \to 2$ the cycle collapses on the {\bf $g = 0$} line. In this regime it shows a non homogeneous running due to the low transient of the trajectory near the GFP, while it  becomes an homogenous slow transient around the NGFP in the limit $d \to d_c$.

Notice that the critical dimension is a function of $n$, $d_c \equiv d_c(n)$, and while for $n = \infty$ the critical dimension is $d_c = 4$, generally holds $d_c(n) < 4$ for a finite value of the parameter $n$. At $d = d_c$ the UV behavior is regulated by a limit cycle whose behavior resembles the one of the Van der Pol oscillator \cite{vander}.
For $d < d_c$ (see left panel of Fig.(\ref{cycles1})) the theory space is now divided in two regions. 
The first is the set of points in parameter space outside the cycle, which trajectories flow towards the UV to the limit cycle and hit in the IR the singularity $\lambda=n/2$ (or flow towards $\lambda = -\infty$). Those are the trajectories which survive for $d>d_c$ and that require higher-order operators in order to cure the IR sector. The second region is the set of points inside the cycle which flow towards it in the UV and towards the NGFP in the IR. The latter case leads to a new interesting scenario in which the UV and IR critical manifolds coincide and the EH truncation is finite at every energy scale. 

For this scenario to be plausible we require the cyclic trajectory to be close enough to the GFP, so that it shows a semiclassical regime. Unfortunately, as can be seen from Fig.(\ref{cycles2}), in the best case ($d_c=4$ for $n=\infty$) a limit cycle with a good semiclassical regime occurs only for $d \approx 3$.
It is also important to stress that the limit cycle never approaches the singularity $\lambda=n/2$, where the EH truncation stops to work. 

Since the Hopf bifurcation is not present in the $\mathbb{R}^d$ projection for the CREH, also for small values of the dimension, we analyzed the behavior of the linearized flow near the NGFP in the case of the full EH truncation, to verify if the Hopf bifurcation is still present in the $S^d$ projection for some value of the parameter $n$.
Numerical results are collected in Table \ref{tab:2} in Appendix A while the $\bet$ functions are listed in Appendix B, and are a simple $d$ dimensional generalization of the results reported in \cite{prop}.
As it can be seen from Table \ref{tab:2} the full theory presents a stable NGFP in the whole $n-d$ plane, which means that the contribution of spin-2 degrees of freedom lower the value of the critical dimension under the  ``critical" value $d = 2$. 

Although such a non trivial behavior in the UV region seems to be a direct consequence of the strong dependence of the flow in the $S^d$ projection on the cutoff parameter $n$, it is interesting to notice that recent investigations based on ``tetrad only" theory spaces  \cite{triat}, and on the  minisuperspace approximation of the EH truncation \cite{cycle3}, also show the presence of limit cycles in the UV and IR limit, respectively.
In the latter case, however, the limit cycle originates by an Hopf bifurcation of a specific cutoff parameter  \cite{priv}, while in our case the bifurcation is governed by the spacetime dimension, so that our limit cycle is UV and not IR.

In closing this section we would like to mention that the intriguing possibility of such a non trivial UV completion in field theory was first pointed out by Wilson in a seminal paper (before the discovery of asymptotic freedom), 
in the context of QCD \cite{wilson71}. In particular it was argued that,  at the experimental level, the presence of a limit cycle would show up in ``perpetual" oscillations in the $e^{+}-e^{-}$ total hadronic cross section in the limit of large momenta.  In the case of gravity the natural arena to discuss this type of phenomenon is the physics of the early Universe, for which an effective Lagrangian ${\cal L}_{eff}(R)$ embodying the properties of the limit cycle
can be determined by using the strategy outlined in \cite{bonanno12}. In the case at hand we expect that  ${\cal L}_{eff}(R) \propto \cos(R/\mu^2)$ where $\mu$ is a renormalization scale.  On the other hand,  discussing the detailed physical implications of this model is beyond the scope of this paper. 

%

%------------------------------------------------------------------------------------------
% SECTION IV: Non polynomial truncations and symmetry breaking  
%------------------------------------------------------------------------------------------
\section{Non polynomial truncations and symmetry breaking}
%------------------------------------------------------------------------------------------
In this section we would like to understand the structure of the UV critical manifold beyond the polynomial truncation discussed in the previous section. 
We are interested in the possibility of having a transition to a phase of broken diffeomorphism invariance at low energy \cite{creh2} and thus we intend to numerically solve (\ref{beh}). In fact although in four dimensions, an ansatz of the type 
\be\label{a1}
V[\chi] = c_1(k)\, \chi^2 + c_2(k)\, \chi^4, \;\;\;\; Z = c_3(k)
\ee 
is an exact polynomial truncation of the coupled Eqs. (\ref{beh}); the question concerning the RG evolution of a generic initial data determined by  (\ref{beh}) 
cannot be answered with this strategy.
The core of the problem can be grasped in the LPA approximation which solves only (\ref{beh1}) by assuming a RG evolution for the wave-function renormalization functional $Z_k$. In fact although the solution of the coupled problem (\ref{beh}) is beyond the aim of this work, we present a successful numerical strategy to deal with (\ref{beh1}) which we hope can eventually be extended to treat the coupled system (\ref{beh1}) and (\ref{beh2}) beyond the simple CREH truncation. In particular, in this section we shall investigate the role played by higher powers of volume operators of the type $\vo=(\int d^d x \sqrt{g})$ in providing a transition to a phase of broken diffeomorphism invariance.
In order to carry out the numerical integration of (\ref{beh1}), it is useful to ``linearize" the evolution equation for the potential by defining the quantity 
\be\label{sing}
W[\chi] = \chi^4\, \left ( 1 +  \frac{V''}{n\,k^2 \,Z\, \chi^2}\right )^{-\gamma}
\ee
with $\gamma = n - 2 > 0$ that diverges at $+\infty$ as the ``spinodal line''  $n\, k^2\, Z_k\, \chi_0^{2\,\nu} +V_k''(\chi_0)=0$ is approached, but it behaves as a power law for large values of the field outside the ``coexistence" region where $n\, k^2\, Z\, \chi_0^2 +V''[\chi_0]<0$.
In terms of this new variable Eq.(\ref{beh1}) reads
\ba\label{nonsing}
&&(2 + \eta)\, n\, k^2\, Z\, \chi^{2}(W^{-\frac{1}{\gamma}}\, \chi^{\frac{4}{\gamma} } - 1)\\[2mm]
&& - n\, k^2\, Z\, \chi^{\frac{4}{\gamma}+2}{\gamma^{-1}}\, W^{-\frac{1}{\gamma}-1}\, k\, \partial_k\, W = A_n\, k^4\, \partial^2_{xx}\,W \, . \nonumber
\ea
The advantage of this manipulation is that Eq. (\ref{nonsing}) is now linear in the second derivative.

Ideally we would like to evolve an initial data defined at the cutoff scale along the RG direction, i.e. towards the infrared.
This is usually achieved by defining the RG time via $k\equiv e^{-t}$ with $t>0$ so that the Cauchy problem is fully determined when $W[\chi,0]$ is fixed and $W[\chi_{\rm out},t]$ is given, $\chi_{\rm out}$ being an asymptotic value of the field ($\chi_{\rm out}\gg 1$ in actual calculations). However, if we intend to do so,  we immediately run into the difficulty that as  $Z<0$, equation (\ref{nonsing}) belongs to the restricted {\it \' elite} of the backward-parabolic equations, i.e. a class of diffusion-type partial differential equation with a {\it negative} diffusion constant. As it is well known, in this case the Cauchy problem is not ``well-posed" (in the sense of \cite{zakh}) and the existence of the solution for generic initial data is not guaranteed even for an infinitesimal time step.

Although we already know that an admissible initial data in four dimensions is precisely of the type (\ref{a1}) it seems that the CREH truncation is one of the few initial data compatible with the general flow equation (\ref{beh}).
In fact, the important question concerning the possibility of developing a non-zero vacuum expectation value of the conformal factor in the  $k \to 0^+$ limit cannot be answered within the simple CREH truncation.

It is therefore necessary  to treat the question as a sort of inverse problem and  to  consider  an integration in the UV direction instead, so that the RG time is defined as $k\equiv e^t$ with $t>0$, the Cauchy problem is well posed, and the solution is unique.
Clearly, once the solution in the deep UV is found it is possible to argue that precisely {\it that solution} is an admissible initial data for a non-singular IR flow. However, also in the case of the UV evolution, due to its strong nonlinearities, a proper numerical strategy is to implement a fully implicit predictor-corrector scheme  on an uniform spatial and temporal grid as discussed in the Appendix C.
\begin{center}
\begin{figure*}
\includegraphics[width=8cm]{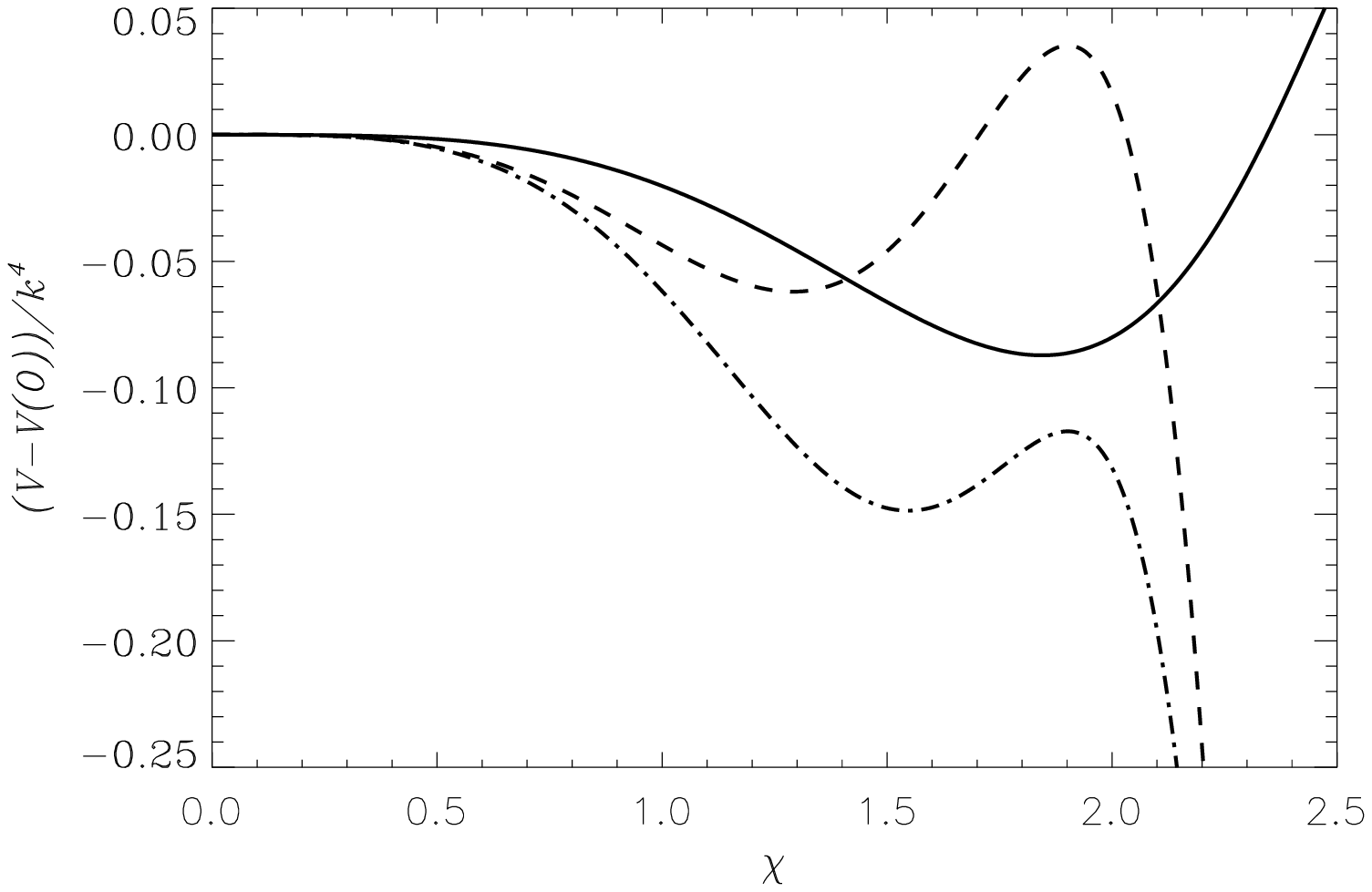}
\includegraphics[width=8cm]{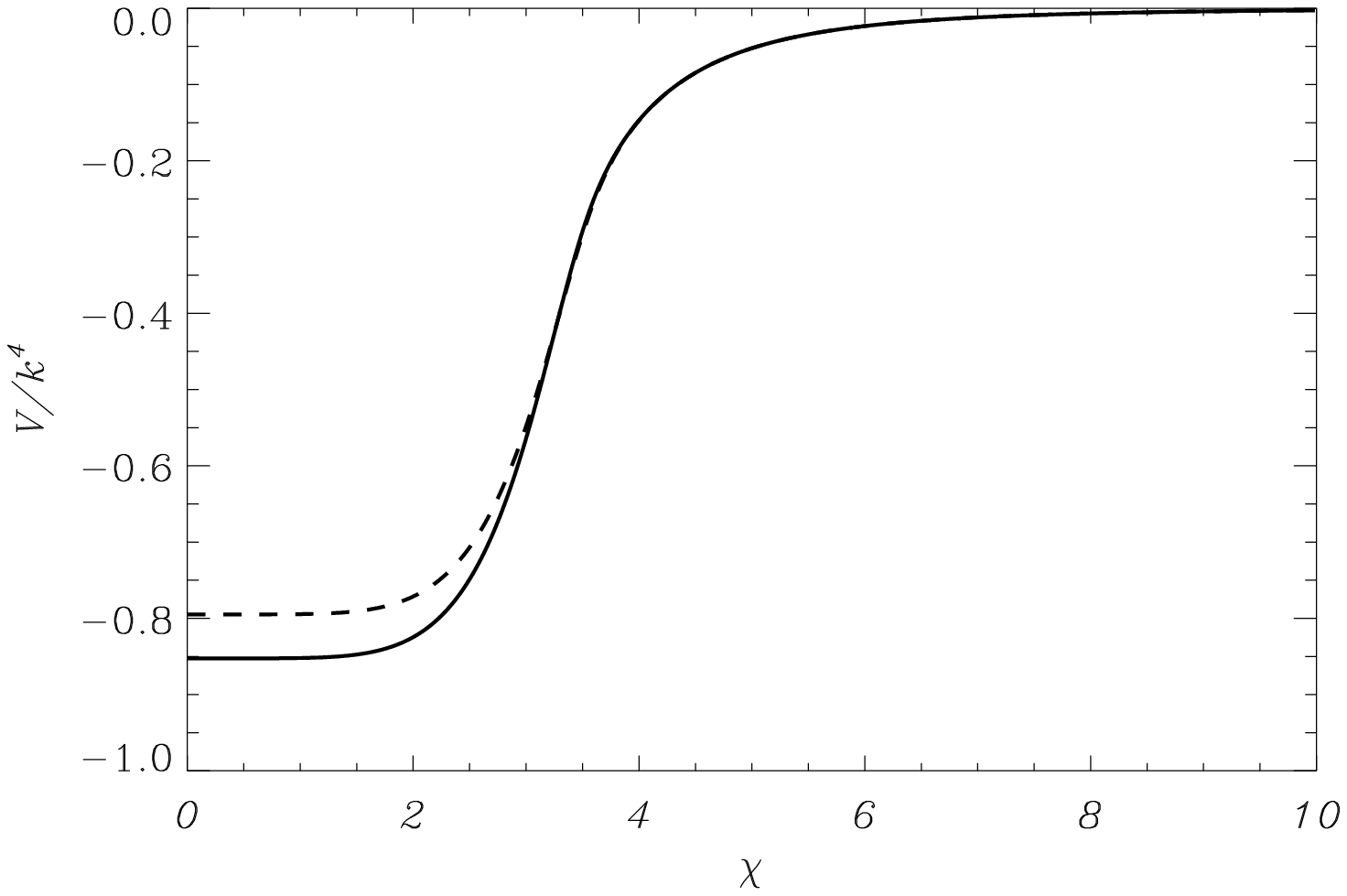}
\caption{The dimensionless initial potential as a function of the RG time $t$ for an initial condition with $\lambda=-1/2$, $\sigma=0.05$ and $\omega=-0.00714$.
Left: the early RG evolution is showed: $t=0$ (dashed line), $t=0.02$, (dotted-dashed line) and $t=0.8$ (solid line). Right: the deep UV fixed point is presented: $t=6$ (solid line) and $t=8$ (dashed line). Further increase in $t$ did not show significant changes in $V$.} 
\label{evol}
\includegraphics[width=8cm]{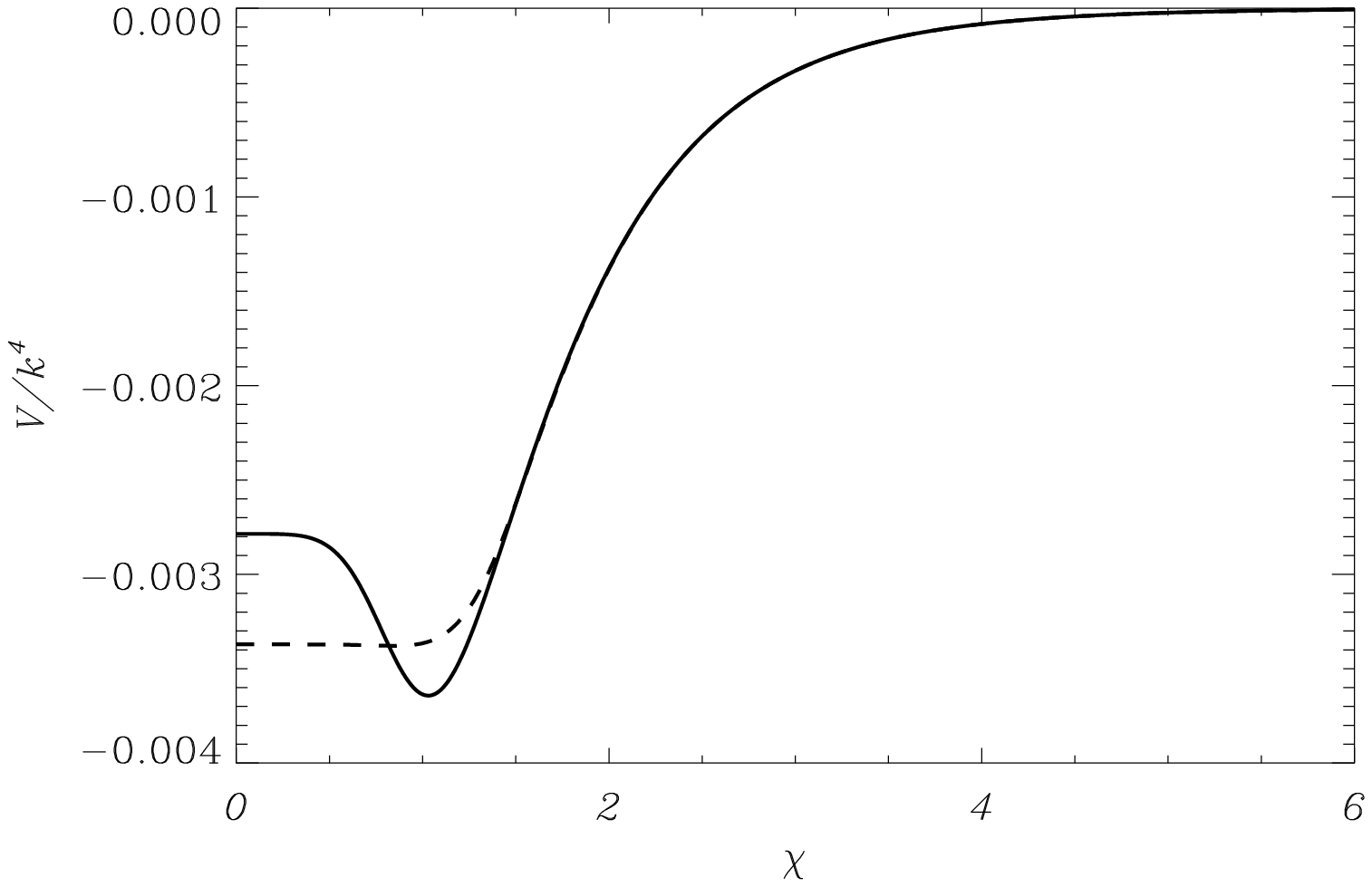}
\includegraphics[width=8cm]{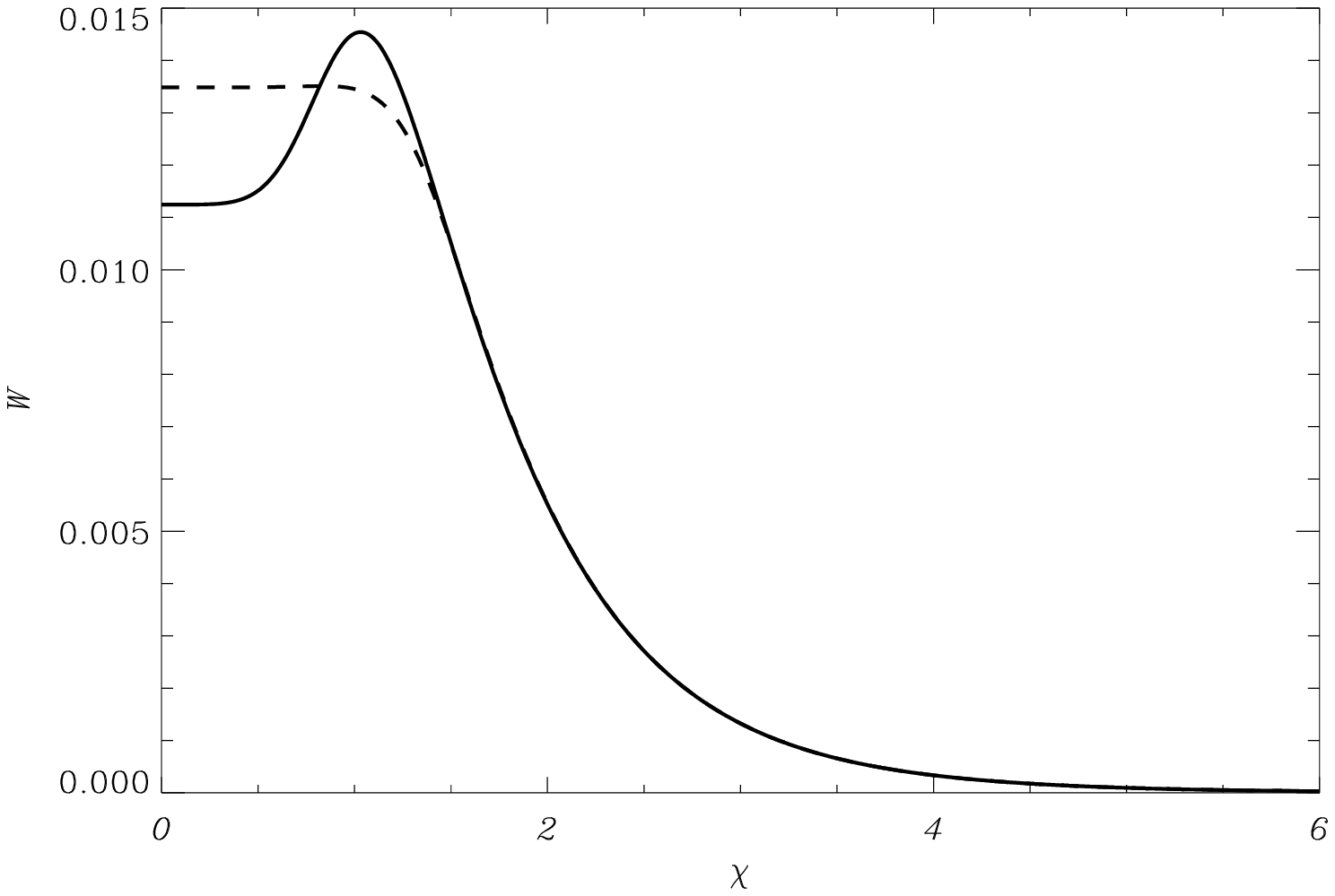}
\caption{Left: dimensionless initial potential as a function of the RG time $t$ for an initial condition with $\lambda=-6$, $\sigma=\omega=-1$ and $n=5$: $t=8$ (solid line), $t=32$ (dashed line). Right: corresponding evolution of the function $W$ is shown.} 
\label{evol2}
\end{figure*}
\end{center}
The boundary condition at $\chi=\chi_{\rm init}$ is of the von Neumann type so that $\partial_x W=0$ at the inner boundary, and $0<\chi_{\rm init}\ll 1$: we have checked that our results are rather insensitive to the choice of  $\chi_{\rm init}$ that could be set arbitrarily close to zero in all calculations (note however that, strictly speaking, $\chi>0$ always). 
The outer boundary is taken at some $ \chi_{\rm out} \gg 1$ where for $W$ a power-law behavior is assumed, like in the more familiar Ising model \cite{bonlac}. The initial value of the potential at the cutoff reads
\be
V[\chi,0] =   \frac{\lambda}{6}\, \chi^4 +  \sigma \chi^6 +\omega \chi^8 \, ,
\ee 
where the bare values of $\lambda$, $\sigma$ and $\omega$ have been chosen in order to display a non-zero minimum as an initial condition.
In addition we have considered also the coupling $\omega<0$ in order to have a real function $W$ for large values of the field. In fact, unless $\omega=\sigma\equiv 0$ no consistent initial condition can be given in all the real line for the potential, as the threshold functions (the denominator in (\ref{beh1}) ) become complex at a finite value of $\chi$ for $\omega>0$.
In solving (\ref{nonsing}) close to the NGFP, we have set $\eta \approx -2$ and $Z\approx -k^2/{g_*}$ in $d=4$ as we are interested in the UV evolution. 

Our results are then summarized in Fig.(\ref{evol}): in the left panel a symmetry breaking initial state evolves towards a convex potential as the UV evolution is followed.
The final, fixed point state, is then reached already for $t=6$ as it can be seen in the right panel. Note the ``flat bottom'' of the potential and the almost exponential suppression at large values of the field in the final solution. We found that the appearance of a ``fixed point'' potential of the type shown in Fig.(\ref{evol}) seems to be quite generic if the initial condition is changed. 

In Fig.(\ref{evol2}) another example of the UV evolution is shown for $n=5$ and for a different set of initial conditions. 
Note in particular that using instead the UV potential at $t=32$ and integrating  towards the IR, a symmetry breaking vacuum appears at low energy. It is convenient to introduce the following order parameter to characterize the phase of the system in this case \cite{polyakov}:
\be
B(L) = \left\langle \exp - \int_0^L g_{\mu\nu} (x (s) {\dot x}^\mu (s) {\dot x}^\nu (s) ds \right\rangle  \, ,
\ee
which is the analogue of the Wilson loops in gauge theory. In our case $\phi_0 =\langle \chi \rangle = const\not = 0$ and $\langle g_{\mu\nu} \rangle = \phi_0\, \delta_{\mu\nu}$ which is therefore a classical flat metric on $\mathbb{R}^4$. For this geometry the well-known heat-kernel behavior at small distances $B(L)\sim L^{-2}$ occurs.

The large field behavior of the potential is characterized by an inverse power behavior for large value of the field thus signaling the presence of non local invariants in the fixed point potential.
The conclusion of our numerical experiment seems to suggest that there exists a more complex fixed point structure, not necessarily of the CREH type, whose precise structure is unaccessible with the more standard $\bet$ function approach.

%----------------------------------------------------------------------------------
%------------------          CONCLUSIONS           --------------------------
%----------------------------------------------------------------------------------
\section{conclusions}
%----------------------------------------------------------------------------------
In this work we further explored the universal properties of the CREH theory around the NGFP. 

The NGFP characterizes the UV evolution for different projection in the theory space and for a large class of threshold functions although we also found the possibility that the continuum limit is defined by means of a limiting-cycle in some cases. 
Moreover it is possible to find an ``optimal'' threshold function for which the renormalization flow minimizes the differences with the calculation of the universal quantities in full Einstein-Hilbert truncation.
 
However, going beyond the CREH truncation has proven to be rather difficult because of the flawed structure of the flow equation in the presence of the conformal factor instability, and only an UV integration was possible. How can we make contact with ``real" gravity that is formally defined at $k = 0$ in more general truncations?
Clearly, a mechanism to stabilize the conformal factor modes is needed and, most probably, the inclusion of the $R^2$ term is essential to provide a consistent truncation \cite{oliverunstable,saur2}. Recent investigations discussed in \cite{dario} for a general $f(R)$ can provide important information for a more complete understanding of the renormalization flow in the theory space.
On the other hand in our investigations we explored the infinite dimensional theory space spanned by the solutions of the (partial differential) flow equation (39) which is not discussed in \cite{dario} because of the technical difficulty in solving the flow equation for a generic $f(R)$ theory.

We find that the structure of the UV region around the NGFP can be richer than expected, with a class of fixed point potentials displaying an inverse power behavior for large value of $\cb$, suggesting the presence of non local volume invariants in the fixed point potential \cite{frank2}.
Within this class of local potentials it seems possible to realize a situation where the diffeomorphism invariance is broken at low energy. It would be interesting to further extend this investigation by including a consistent running of $Z$ and the contribution from the $R^2$ term to see the infrared flow structure of an $R^2$-stabilized conformal factor beyond polynomial truncations, and we hope to address this point in a following work.

%---------------------------------
%    Acknowledgements
%---------------------------------
\begin{acknowledgements}
%---------------------------------
We would like to thank Dario Benedetti and  Roberto Percacci for important comments and discussions.  One of us (F.G.) is grateful to INFN, Rome  University ``La Sapienza",  and INAF for financial support.
\end{acknowledgements}

	\clearpage

\begin{widetext}
%---------------------------------
%          APPENDIX
%---------------------------------
\appendix
%---------------------------------
\section{CRITICAL EXPONENTS AND UNIVERSAL QUANTITIES}
In this appendix, we show Tables I and II.
\begin{table*}[!htbp]
\caption{The fixed point values and the critical exponents obtained in four dimensions from the $\bet$ functions (\ref{b1}) and (\ref{b2}) for various values of the cutoff parameter $n$ compared with those previously obtained from the full EH gravity in \cite{prop}, on the right.
\label{tab:1}}
\begin{center}
\begin{tabular}{c|ccccc|ccccc|ccccc}
\hline\hline
&  & & \multicolumn{2}{c}{CREH-$S^4$} &  \multicolumn{6}{c}{CREH-$\mathbb{R}^4$} &\multicolumn{4}{c}{Full EH-$S^4$}  \\
\hline
{\bf n}&  $\las$  &    $\gas$    & $\las\gas$ & $\tp$   &  $\tpp$ &$\las$ & $\gas$& $\las\gas$&$\tp$  & $\tpp$ &  $\las$   & $\gas$  & $\las\gas$ &   $\tp$   &   $\tpp$\\  
\hline
3     &    1.125  &    1.571     &    1.767   &  3      &   4.795 & 0.800 & 2.084&1.670 & 8.580 & 0	 &  0.355   &  0.388   &   0.138   &   1.835   &    1.300\\
4     &    1.2    &    1.810     &    2.171   &  1.5    &   4.213 & 0.837 & 2.666&2.234 & 5.721 & 2.928  &  0.265   &  0.472   &   0.125   &   1.770   &    1.081\\
5     &    1.25   &    1.885     &    2.356   &  1	&   3.873 & 0.867 & 2.914&2.528 & 5.000 & 3.428  &  0.230   &  0.517   &   0.119   &   1.750   &    1.000\\
6     &    1.285  &    1.914     &    2.461   &  0.75   &   3.665 & 0.889 & 3.041&2.706 & 4.578 & 3.627  &  0.211   &  0.546   &   0.115   &   1.742   &    0.959\\
8     &    1.333  &    1.930     &    2.574   &  0.5	&   3.427 & 0.921 & 3.159&2.910 & 4.102 & 3.788  &  0.191   &  0.582   &   0.111   &   1.734   &    0.916\\
10    &    1.364  &    1.931     &    2.633   &  0.375  &   3.295 & 0.941 & 3.209&3.023 & 3.839 & 3.850  &  0.181   &  0.603   &   0.109   &   1.731   &    0.894\\
15    &    1.406  &    1.923     &    2.703   &  0.230  &   3.129 & 0.971 & 3.253&3.161 & 3.511 & 3.903  &  0.169   &  0.630   &   0.106   &   1.727   &    0.868\\
20    &    1.428  &    1.914     &    2.735   &  0.167  &   3.050 & 0.987 & 3.265&3.225 & 3.356 & 3.919  &  0.163   &  0.644   &   0.105   &   1.725   &    0.856\\
30    &    1.451  &    1.904     &    2.764   &  0.107  &   2.974 & 1.004 & 3.271&3.285 & 3.206 & 3.929  &  0.158   &  0.658   &   0.104   &   1.723   &    0.846\\
50    &    1.470  &    1.894     &    2.785   &  0.062  &   2.914 & 1.018 & 3.271&3.331 & 3.089 & 3.933  &  0.154   &  0.668   &   0.103   &   1.722   &    0.837\\
100   &    1.485  &    1.886     &    2.800   &  0.030  &   2.871 & 1.028 & 3.269&3.364 & 3.003 & 3.934  &  0.152   &  0.676   &   0.102   &   1.721   &    0.831\\
300   &    1.495  &    1.880     &    2.810   &  0.010  &   2.842 & 1.036 & 3.266&3.385 & 2.948 & 3.934  &  0.150   &  0.682   &   0.102   &   1.721   &    0.828\\
$\infty$ & 1.5    &    1.880     &    2.815   &  0      &   2.820 & 1.040 & 3.265&3.396 & 2.920 & 3.923  &  0.103   &  0.685   &   0.070   &   1.720   &    0.826\\
\hline\hline
\end{tabular}
\end{center}
\end{table*}
\begin{table*}[!htbp]
\caption{The fixed point values and the critical exponents obtained for the cutoff parameter $n = 4$ from the $\bet$ functions (\ref{b1}) and (\ref{b2}) for various values of the dimension compared with those obtained from the full EH gravity, on the right.
\label{tab:2}}
\begin{center}
\begin{tabular}{c|ccccc|ccccc|ccccc}
\hline\hline
&  & & \multicolumn{2}{c}{CREH-$S^4$} &  \multicolumn{6}{c}{CREH-$\mathbb{R}^4$} &\multicolumn{4}{c}{Full EH-$S^4$}  \\
\hline
{\bf d}&  $\las$  &    $\gas$    & $\las\gas$ & $\tp$   &  $\tpp$ &$\las$ & $\gas$& $\las\gas$&$\tp$  & $\tpp$ &  $\las$   & $\gas$  & $\las\gas$ &   $\tp$   &   $\tpp$\\  
\hline
2     	&    0  	   &    0    	 	&    0   		&  $- \infty$   	&   0 		& 0	 	& 0		& 0 		& 0 		& 0		& 0         	 &  0	       	&   0  	 	&   2   	  &    0\\
2.2    &    0.1825  &    4.07E-4    &    7.44E-5 	&  -38.056    	&   0 		& 0.0381 	& 0.0214	& 8.18E-4	& 0.8153 	& 0.9756  	& -0.0176  &  0.0184 & -3.25E-4  	&   2.0025   &    0\\
2.4    &    0.3381  &    5.24E-3    &    1.77E-3 	&  -17.356	    	&   0 		& 0.0968 	& 0.0632	& 6.12E-3 & 1.1495 	& 1.5524  	& -0.0277  &  0.0428 &  -1.19E-3  	&   1.9882   &    0\\
2.6    &    0.4748  &    0.0223	&    0.0106  	&  -9.8196    	&   0 		& 0.1676 	& 0.1318	& 0.0220 	& 1.5641 	& 1.9715	& -0.0291  &  0.0742 & -2.16E-3   	&   1.9552   &    0\\
2.8    &    0.5980  &    0.0619     	&    0.0370 	&  -5.1014		&   0 		& 0.2474 	& 0.2369	& 0.0586	& 2.0317	& 2.3065  	& -0.0204  &  0.1132 &  -2.31E-3   	&   1.8981   &    0\\
3    	&    0.7111  &    0.1368    	&    0.0973 	&  -2.  		&   2.2360 & 0.3344 	& 0.3917	& 0.1309 	& 2.5433	& 2.5771  	&  0   	 &  0.16  	 &  0  		&   1.8   	  &    0\\
3.2    &    0.8168  &    0.2637     	&    0.2154   	&  -1.05  		&   3.1098 & 0.4270 	& 0.6127	& 0.2616 	& 3.0957	& 2.7887  	&  0.0331 	 &  0.2140 &  7.09E-3   	&   1.5031   &    0\\
3.4    &    0.9171  &    0.4639     	&    0.4254   	&  -0.2832 	&   3.5584 & 0.5243 	& 0.9210	& 0.4829	& 3.6884 	& 2.9400  	&  0.0791 	 &  0.2737 &  0.0216   	&   1.5093   &    0.4259\\
3.6    &    1.0136  &    0.7643     	&    0.7747   	&  0.3727  	&   3.8450 & 0.6256 	& 1.3431	& 0.8402 	& 4.3220	& 3.0241 	&  0.1358   &  0.3373 &  0.0458   	&   1.5738   &    0.6480\\
3.8    &    1.1075  &    1.1989     	&    1.3278   	&  0.9587  	&   4.0516 & 0.7303 	& 1.9115	& 1.3959 	& 4.9986	& 3.0278  	&  0.1994   &  0.4035 &  0.0804   	&   1.6579   &    0.8617\\
4    	&    1.2  	   &    1.8095    	&    2.1714   	&  1.5  		&   4.2130 & 0.8379 	& 2.6660	& 2.2341 	& 5.7211	& 2.9284  	&  0.2653   &  0.4724 &  0.1253  	&   1.7690   &    1.0810\\
4.2   	&    1.2918  &    2.6471    	&    3.4197   	&  2.0134  	&   4.3464 & 0.9483 	& 3.6549	& 3.4660 	& 6.4931	& 2.6839  	&  0.3299   &  0.5452 &  0.1799 	&   1.9096   &    1.3047\\
4.4   	&    1.3839  &    3.7724     	&    5.2209   	&  2.5111  	&   4.4608 & 1.0611 	& 4.9352	& 5.2368 	& 7.3194	& 2.2018  	&  0.3919   &  0.6238 &  0.2444  	&   2.0789   &    1.5288\\
4.6   	&    1.4771  &    5.2560     	&    7.7639   	&  3.0027      	&   4.5609 & 1.1762 	& 6.5736	& 7.7321 	& 8.2057	& 1.1214  	&  0.4508   &  0.7096 &  0.3199  	&   2.2746   &    1.7504\\
4.8   	&    1.5721  &    7.1786     	&    11.285  	&  3.4964      	&   4.6491 & 1.2935 	& 8.6465	& 11.185 	& 11.129	& 0  		&  0.5074   &  0.8036 &  0.4078 	&   2.4945   &    1.9684\\
\hline\hline
\end{tabular}
\end{center}
\end{table*}

	\clearpage

\section{$\bet$ FUNCTIONS}
In this appendix are listed the explicit expressions of the $\bet$ functions in $d$ dimensions for different choices of the cutoff function. The $\bet$ functions obtained using the cutoff (\ref{3}) will be referred as smooth proper-time cutoff, where the limits $n \to d/2$ and $n \to \infty$ will be called respectively sharp momentum cutoff and sharp proper-time cutoff.
\subsection{The projection on $S^d$}
\begin{itemize}

\item{CREH - smooth proper-time cutoff

\begin{align}
\beta_g = g_k\, \left( d - 2 -\frac{2^{2-d}\, (d-2)\, \pi^{1 - \frac{d}{2}}\, g_k\, n^n\, \Gamma \left(-\frac{d}{2} + n + 1 \right)}{\left(n - \frac{d \left(\frac{2 d}{d-2} - 1\right)\, \lambda_k }{2\, (d-1)}\right)^{n - \frac{d}{2} + 1}\, (d-1)\, \Gamma(n)} \right) \, ,
\end{align}

\begin{align}
\beta_\lambda = \frac{2^{3-d}\, \pi^{1-\frac{d}{2}}\, g_k\, n^n\, \Gamma \left( n - \frac{d}{2}\right ) }{\Gamma(n)\,  \left(n - \frac{d\, \left(\frac{2\, d}{d-2} - 1\right)
   \lambda_k }{2\, (d-1)}\right)^{n - \frac{d}{2}}}  + \lambda  \left(-2 -\frac{2^{2-d}\, (d-2)\, \pi^{1 - \frac{d}{2}}\, g_k\, n^n\, \Gamma \left(-\frac{d}{2} + n + 1 \right)}{\left(n - \frac{d \left(\frac{2
   d}{d-2} - 1\right)\, \lambda_k }{2\, (d-1)}\right)^{n - \frac{d}{2} + 1}\, (d-1)\, \Gamma(n)}\right) \, .
\end{align}}

\vspace{2em}

\item{CREH - sharp proper-time cutoff

\begin{align}
\beta_g = g_k \,\left(d - 2 - \frac{2^{2-d}\, (d-2)\, \pi^{1-\frac{d}{2}}\, g_k\, }{d-1} \, \exp \left [\frac{d\, \left(\frac{2 \,d}{d-2} - 1\right)\, \lambda_k }{2 (d-1)} \right ] \right) \, ,
\end{align}

\begin{align}
\beta_{\lambda}= 2^{3-d}\, \pi^{1-\frac{d}{2}}\, g_k\, \exp \left[\frac{d\, \left(\frac{2\, d}{d-2} - 1\right)\, \lambda_k }{2\, (d-1)} \right] + \lambda_k\,  \left( - 2 - \frac{2^{2-d}\, (d-2)\, \pi^{1-\frac{d}{2}}\, g_k\, }{d-1} \,
\exp \left [\frac{d\, \left(\frac{2 \,d}{d-2} - 1\right)\, \lambda_k }{2 (d-1)} \right ] \right) \, .
\end{align}}

\vspace{2em}

\item{CREH - sharp momentum cutoff

\begin{align}
\beta_g = g_k\, \left( d - 2 +\frac{2^{3 - \frac{3\, d}{2}}\, (d-2)\, \pi^{1 - \frac{d}{2}}\, d^{d/2}\, g_k}{(d-1)\, \left(\frac{d \,\left(\frac{2\, d}{d - 2} - 1 \right)\, \lambda_k }{d - 1} - d \right) \Gamma
   \left(\frac{d}{2} \right)}\right)\, ,
\end{align}

\begin{align}
\beta_\lambda=
\left(- 2 + \frac{2^{3 - \frac{3\, d}{2}}\, (d-2)\, \pi^{1 - \frac{d}{2}}\, d^{d/2}\, g_k}{(d-1)\, \left(\frac{d \,\left(\frac{2\, d}{d - 2} - 1 \right)\, \lambda_k }{d - 1} - d \right) \Gamma
   \left(\frac{d}{2} \right)} \right)\, \lambda_k  - \frac{2^{3 - \frac{3\, d}{2}}\, d^{d/2}\, \pi^{1 - \frac{d}{2}}\, g_k\, \log \left[1 - \frac{\left(\frac{2\, d}{d-2} - 1 \right)\, \lambda_k }{d-1} \right]}{\Gamma
   \left(\frac{d}{2}\right)} \, .
\end{align}}

\vspace{2em}

\item{FULL EH - smooth proper-time cutoff

\begin{align}
\beta_g = g_k\, \left(  d - 2 + \frac{2^{2-d}\, \pi^{1-\frac{d}{2}}\, g_k\, \left( - d\, (5\, d - 7)\, n^{n+1}\, (n - 2\, \lambda_k)^{\frac{d}{2} - n - 1} - 4\, ( d + 6 )\, n^{d/2}\right) \Gamma
   \left( - \frac{d}{2} + n + 1 \right)}{3\, \Gamma (n+1)} \right) \, ,
\end{align}
\begin{align}
\beta_\lambda= \frac{2^{2-d}\, d\, \pi^{1-\frac{d}{2}}\, g_k\, (n - 2\, \lambda_k )^{-n}\, \left((d+1)\, n^n\, (n - 2\, \lambda_k )^{d/2}- 4\, n^{d/2}\, (n - 2\, \lambda_k )^n \right) \Gamma
   \left(n - \frac{d}{2} \right)}{\Gamma(n) } +\nonumber
\end{align}
\begin{align}
    + \left(- 2 + \frac{2^{2-d}\, \pi^{1-\frac{d}{2}}\, g_k\, \left( - d\, (5\, d - 7)\, n^{n+1}\, (n - 2\, \lambda_k)^{\frac{d}{2} - n - 1} - 4\, ( d + 6 )\, n^{d/2}\right) \Gamma
   \left( - \frac{d}{2} + n + 1 \right)}{3\, \Gamma (n+1)}\right)\, \lambda_k \, .
\end{align}}

\vspace{2em}

\item{FULL EH - sharp proper-time cutoff

\begin{align}
\beta_g = g_k \left(d - 2 - \frac{1}{3}\, 2^{2-d}\, \pi^{1-\frac{d}{2}}\, g_k\, \left(d\, (5\, d - 7)\, \exp \left[2\, \lambda_k \right] + 4\, (d + 6) \right) \right) \, ,
\end{align}
\begin{align}
\beta_\lambda = 2^{2-d}\, d\, \pi^{1 - \frac{d}{2}}\, g_k\, \left( (d+1)\, \exp \left[ 2\, \lambda_k \right] - 4 \right)  + \left ( -  2 - \frac{1}{3}\, 2^{2-d}\, \pi^{1-\frac{d}{2}}\, g_k\, (d\, (5\, d - 7)\, \exp \left[2\, \lambda_k \right] + 4\, (d + 6) \right) \lambda_k \, .
\end{align}}

\vspace{2em}

\item{FULL EH - sharp momentum cutoff

\begin{align}
\beta_g = g_k \left(d - 2 - \frac{2^{3-d}\, d^{d/2}\, \pi^{1-\frac{d}{2}}\, 2^{-d/2}\, g_k\, \left(3\, d\, ((d-1)\, d + 4) - 2\, \left(d^2 + d + 24\right)\, \lambda_k \right)}{3\, (d - 4 \, \lambda_k )\, \Gamma \left(\frac{d}{2} + 1\right)}\right) \, ,
\end{align}
\begin{align}
\beta_\lambda= \left( - 2 - \frac{2^{3-d}\, d^{d/2}\, \pi^{1-\frac{d}{2}}\, 2^{-d/2}\, g_k\, \left(3\, d\, ((d-1)\, d + 4) - 2\, \left(d^2 + d + 24\right)\, \lambda_k \right)}{3\, (d - 4 \, \lambda_k )\, \Gamma \left(\frac{d}{2} + 1\right)} \right)\, \lambda_k + \nonumber
\end{align}
\begin{align}
+ \frac{- 2^{2-d}\, (d+1)\, \pi^{1 - \frac{d}{2}}\, 2^{-d/2} \,d^{\frac{d}{2}+1}\, g_k\, \log \left[1- \frac{4}{d}\, \lambda_k \right]}{\Gamma \left(\frac{d}{2}\right)} \, .
\end{align}}

\subsection{The projection on $\mathbb{R}^d$}

\item{CREH - smooth proper-time cutoff

\begin{align}
\beta_g = g_k\, \left(d - 2 - \frac{2^{2-d}\, d^2\, (d + 2)^2\, \pi^{1-\frac{d}{2}}\, g_k\, \lambda_k^2\, n^n\, \Gamma \left(-\frac{d}{2} + n + 3 \right)}{3\, (d - 2)^3\, (d - 1)^3\, \Gamma (n)\, \left( n - \frac{d\, \left(\frac{2\, d}{d-2} - 1 \right)\, \lambda_k }{2\, (d-1)}\right)^{\frac{1}{2} (- d + 2\, n + 6)}}  \right) \, ,
\end{align}
\begin{align}
\beta_\lambda = \frac{2^{3-d}\, \pi^{1 - \frac{d}{2}}\, g_k\, \Gamma \left(n - \frac{d}{2} \right)}{\Gamma (n)\, \left(1 - \frac{d\, (d+2)\, \lambda_k }{2\, (d-2)\, (d - 1)}\right)^{\frac{1}{2} (2\,n - d)}} + 
\left(- 2 - \frac{2^{2-d}\, d^2\, (d + 2)^2\, \pi^{1-\frac{d}{2}}\, g_k\, \lambda_k^2\, n^n\, \Gamma \left(-\frac{d}{2} + n + 3 \right)}{3\, (d - 2)^3\, (d - 1)^3\, \Gamma (n)\, \left( n - \frac{d\, \left(\frac{2\, d}{d-2} - 1 \right)\, \lambda_k }{2\, (d-1)}\right)^{\frac{1}{2} (- d + 2\, n + 6)}}\right) \lambda_k \, .
\end{align}}

\vspace{2em}

\item{CREH - sharp proper-time cutoff

\begin{align}
\beta_g = g_k\, \left(d - 2 - \frac{2^{2-d}\, d^2 \,(d + 2)^2\, \pi^{1 - \frac{d}{2}}\, g_k\, \lambda_k^2}{3\, (d-2)^3 \,(d-1)^3}\, \exp \left[ \frac{d\, \left(\frac{2\, d}{d - 2} - 1 \right)\, \lambda_k }{2\, (d-1)} \right]  \right) \, ,
\end{align}
\begin{align}
\beta_\lambda = 2^{3-d}\, \pi^{1 - \frac{d}{2}}\, g_k\, \exp \left[ \frac{d\, (d+2)\, \lambda_k }{2\, (d - 2)\, (d - 1)} \right]  + \left(- 2 -  \frac{2^{2-d}\, d^2 \,(d + 2)^2\, \pi^{1 - \frac{d}{2}}\, g_k\, \lambda_k^2}{3\, (d-2)^3 \,(d-1)^3}\, \exp \left[ \frac{d\, \left(\frac{2\, d}{d - 2} - 1 \right)\, \lambda_k }{2\, (d-1)} \right]  \right) \, \lambda_k\, .
\end{align}}

\vspace{2em}

\item{CREH - sharp momentum cutoff

\begin{align}
\beta_g = g_k \left(d - 2 -\frac{2^{5-\frac{3\, d}{2}}\, (d + 2)^3\, \pi^{1 - \frac{d}{2}}\, d^{d/2}\, g_k\, \lambda_k^3\, \left(3\, \left(d^2 - 3 \,d + 2 \right)^2 + ( d + 2 )^2\, \lambda_k^2 - 3\, (d-2)\, (d-1)\, (d+2) \,\lambda_k \right)}{3 (d-2)^3 (d-1)^3 \left(d^2-d (\lambda +3)-2 \lambda +2\right)^3 \Gamma \left(\frac{d}{2}+1\right)}\right) \, ,\nonumber\\
\end{align}
\begin{align}
\beta_\lambda =  \lambda_k\,  \left(- 2 - \frac{2^{5 - \frac{3\, d}{2}}\, (d + 2)^3\, \pi^{ 1 - \frac{d}{2}}\, d^{d/2}\, g_k\, \lambda_k^3\, \left(3\, \left(d^2 - 3\, d + 2 \right)^2 + ( d + 2 )^2\,
   \lambda_k^2 - 3\, (d-2)\, (d-1)\, (d+2)\, \lambda_k \right)}{3\, (d-2)^3\, (d-1)^3\, \left(d^2 - d\, (\lambda_k +3) - 2 \lambda_k +2 \right)^3\, \Gamma \left(\frac{d}{2}+1\right)}\right) + \nonumber
\end{align}
\begin{align}
- \frac{ 2^{3 - \frac{3\, d}{2}}\, \pi^{1 - \frac{d}{2}}\, d^{d/2}\, g_k\, \log \left[1-\frac{(d + 2)\, \lambda_k}{(d-2)\, (d-1)}\right]}{\Gamma \left(\frac{d}{2}\right)} \, .
\end{align}}

\end{itemize}

\section{NUMERICAL STRATEGY}
In solving the flow equation, the predictor step is computed at times  $t=(j+1/2)\, \Delta t$ so that we can discretize (\ref{nonsing}) according to the scheme
\ba\label{pred}
&&\frac{1}{h^2}\, \delta^2_x\, W_{i,j+1/2} = \\[2mm]
&&\frac{n \, Z_{i,j}}{A_n}\, (2 + \eta)\, e^{- 2 \,t }\,(ih)^2\, \left ( W^{-\frac{1}{\gamma}}_{i,j}(ih)^{\frac{4}{\gamma}} - 1 \right )
- \frac{n \, Z_{i,j}}{A_n\, \gamma}\, e^{ - 2 \, t}(ih)^{\frac{4}{\gamma} + 2}\, W^{-\frac{1}{\gamma} - 1}_{i,j}\, \frac{2}{q}\, \left ( W_{i,j + 1/2} - W_{i,j} \right )    \, , \nonumber
\ea
being $h=1/\Delta \chi$ the spatial grid spacing, $q=1/\Delta t$ the temporal grid spacing and, as usual, $\delta^2_x\, W_i = W_{i-1}-2 W_i +W_{i+1}$. 
The corrector step is instead given by 
\ba\label{corr}
&&\frac{1}{2\, h^2}\,\delta^2_x\, \left [W_{i,j+1}+W_{i,j}\right ] = \\[2mm]
&&\frac{n \, Z_{i,j+1/2}}{A_n}\, (2 + \eta)\, e^{- 2 \,t}(ih)^2\, \left ( W^{-\frac{1}{\gamma}}_{i,j+1/2}(ih)^{\frac{4}{\gamma}} -1\right ) 
-\frac{n \, Z_{i,j+1/2}}{A_n\, \gamma}\, e^{- 2 \,t}(ih)^{\frac{4}{\gamma} + 2}\, W^{-\frac{1}{\gamma}-1}_{i,j+1/2}\, \frac{1}{q}\, \left ( W_{i,j+1}-W_{i,j} \right )  \, , \nonumber
\ea
and the solution at $j+1/2$ in (\ref{corr}) is obtained from (\ref{pred}) from the  solution of the linear tridiagonal system problem in the predictor step.
As a consequence  (\ref{corr}) also reduces to a  linear problem for the $j+1$ time step that can be conveniently solved by standard tridiagonal solvers. The method is thus unconditionally stable and $O[h^2+q^2]$ accurate \cite{ames}.
\end{widetext}

%----------------------------------
%       BIBLIOGRAPHY
%----------------------------------


\begin{thebibliography}{99} 
%\bibitem{sav}
%G.K. Savvidy, Phys. Lett. B 71 (1977), 133-134


%Kiefer:2007ria
\bibitem{kiefer}
For general introductions see C.~Kiefer, \textit{Quantum Gravity}, Second Edition, Oxford Science Publications, Oxford (2007);
H.~Hamber,  \emph{Quantum Gravitation}, Springer, Berlin (2008).
%

%Ashtekar:1991hf
\bibitem{A}
A.~Ashtekar, \textit{Lectures on non-perturbative canonical gravity}, World Scientific, Singapore (1991); 
A.~Ashtekar and J.~Lewandowski, Class.\ Quant.\ Grav.\ 21 (2004) R53.
%

%Rovelli:2004tv
\bibitem{R}
C.~Rovelli, \textit{Quantum Gravity}, Cambridge University Press, Cambridge (2004).
%

%Thiemann:2007zz
\bibitem{T}
Th.~Thiemann, \textit{Modern Canonical Quantum General Relativity}, Cambridge University Press, Cambridge (2007).
%


%Weinberg:1980gg
\bibitem{wein}
S.~Weinberg in \textit{General Relativity, an Einstein Centenary Survey}, S.W.~Hawking and W.~Israel (Eds.), Cambridge University Press (1979);
S.~Weinberg, \mbox{ arXiv:0903.0568 [hep-th]}. 



%Reuter:1996cp
\bibitem{mr}
M.~Reuter, Phys.\ Rev.\ D 57 (1998) 971 and \mbox{hep-th/9605030}.


%Dou:1997fg
\bibitem{percadou}
D.~Dou and R.~Percacci, Class.\ Quant.\ Grav. 15 (1998) 3449.
%


%Lauscher:2001ya
\bibitem{oliver1}
O.~Lauscher and M.~Reuter, Phys.\ Rev.\ D 65 (2002) 025013 and 
\mbox{hep-th/0108040.}


%Reuter:2001ag
\bibitem{frank1}
M.~Reuter and F.~Saueressig, Phys.\ Rev.\ D 65 (2002) 065016 and \mbox{hep-th/0110054.}


%Lauscher:2002sq
\bibitem{oliver2}
O.~Lauscher and M.~Reuter, Phys.\ Rev.\ D 66 (2002) 025026 and \mbox{hep-th/0205062.}



%Lauscher:2001rz
\bibitem{oliver3}
O.~Lauscher and M.~Reuter, Class.\ Quant.\ Grav.19 (2002) 483 and \mbox{hep-th/0110021.}



%Lauscher:2001cq
\bibitem{oliver4}
O.~Lauscher and M.~Reuter, Int.J.\ Mod.\ Phys.A 17 (2002) 993 and \mbox{hep-th/0112089.}




%Souma:1999at
\bibitem{souma}
W.~Souma, Prog.\ Theor.\ Phys.\ 102 (1999) 181.




%-------------------------------------------------------------------------------------------------------
\bibitem{frank2}
M.~Reuter and F.~Saueressig, 
Phys.\ Rev.\ D 66 (2002) 125001 and \mbox{hep-th/0206145;}
Fortschr.\ Phys. 52 (2004) 650 and \mbox{hep-th/0311056.}





%Bonanno:2004sy
\bibitem{prop}
A.~Bonanno and M.~Reuter,
JHEP 02 (2005) 035 and \mbox{hep-th/0410191.}



%-------------------------------------------------------------------------------------------------------
\bibitem{oliverbook}
For reviews see: 
M.~Reuter and F.~Saueressig, \mbox{arXiv:0708.1317 [hep-th]}; O.~Lauscher and M.~Reuter in \textit{Quantum Gravity}, B.~Fauser, 
J.~Tolksdorf and E.~Zeidler (Eds.), Birkh\"auser, Basel (2007) and \mbox{hep-th/0511260}; 
O.~Lauscher and M.~Reuter in \textit{Approaches to Fundamental Physics},  
I.-O.~Stamatescu and E.~Seiler (Eds.), Springer, Berlin (2007).


%-------------------------------------------------------------------------------------------------------
\bibitem{perper1}
R.~Percacci and D.~Perini,
Phys.\ Rev.\ D 67 (2003) 081503; 
Phys.\ Rev.\ D 68 (2003) 044018;
Class.\ Quant.\ Grav.21 (2004) 5035.

%-------------------------------------------------------------------------------------------------------
\bibitem{codello}
A.~Codello and R.~Percacci,
Phys.\ Rev.\ Lett.97 (2006) 221301; 
A.~Codello, R.~Percacci and C.~Rahmede,
Int.J.\ Mod.\ Phys.A23 (2008);  preprint \mbox{arXiv:0805.2909 [hep-th].}


%-------------------------------------------------------------------------------------------------------
\bibitem{litimgrav}
D.~Litim,
Phys.\ Rev.\ Lett.92 (2004) 201301;AIP Conf.Proc.841 (2006) 322;
P.~Fischer and D.~Litim, Phys.\ Lett.\ B 638 (2006) 497;
AIP Conf.Proc.861 (2006) 336.



%Machado:2007ea
\bibitem{frankmach}
P.~Machado and F.~Saueressig, Phys.\ Rev. D 77 (2008) 124045.



%--------------------------------------------------------------------------------------------------------
%BMS1 BMS2
\bibitem{BMS}
D.~Benedetti, P.~Machado and F.~Saueressig,
Mod.Phys.Lett.A24 (2009) 2233 \mbox{arXiv:0901.2984 [hep-th]} and Nucl.Phys.B824 (2010) 168 \mbox{arXiv:0902.4630 [hep-th].}



%Lauscher:2005qz
\bibitem{oliverfrac}
O.~Lauscher and M.~Reuter, JHEP 10 (2005) 050 and \mbox{hep-th/0508202.}


%Reuter:2005bb
\bibitem{jan1}
M.~Reuter and J.-M.~Schwindt, JHEP 01 (2006) 070 and \mbox{hep-th/0511021.}


%Reuter:2006zq
\bibitem{jan2}
M.~Reuter and J.-M.~Schwindt, JHEP 01 (2007) 049 and \mbox{hep-th/0611294.}


%--------------------------------------------------------------------------------------------------------
\bibitem{max}
P.~Forg\'acs and M.~Niedermaier, \mbox{hep-th/0207028; } M.~Niedermaier, JHEP 12 (2002) 066;
Nucl.\ Phys.\ B 673 (2003) 131; Class.\ Quant.\ Grav.24 (2007) R171.


%--------------------------------------------------------------------------------------------------------
\bibitem{livrev}
For detailed reviews of asymptotic safety in gravity see: M.~Niedermaier and M.~Reuter, Living Reviews in Relativity 9 (2006) 5; R.~Percacci, \mbox{arXiv:0709.3851 [hep-th].}


%Nagy:2012rn
\bibitem{nagi}
S.\, Nagy, J.\, Krizsan, K.\, Sailer, \mbox{arXiv:1203.6564 [hep-th].}



%Abbott:1981ke
\bibitem{abbott}
L. F. Abbott, Acta Physica Polonica, Vol. B13 (1982).



%
%--------------------------------------------------------------------------------------------------------
%Reuter:2008wj Reuter:2009kq
\bibitem{creh1}
M.~Reuter and H.~Weyer, Phys.\ Rev.\  D {\bf 79} (2009) 105005 and \mbox{arXiv:0801.3287 [hep-th]};
 Gen.\ Rel.\ Grav.\  {\bf 41} (2009) 983 and \mbox{ arXiv:0903.2971 [hep-th]}.


%Reuter:2008qx
\bibitem{creh2}
M.~Reuter and H.~Weyer, Phys.\ Rev.\  D {\bf 80} (2009) 025001, and \mbox{arXiv:0804.1475 [hep-th]}.
%



%--------------------------------------------------------------------------------------------------------
% Manrique:2009uh Manrique:2010mq
\bibitem{elisa}
E. Manrique and M. Reuter, Annals Phys. 325 (2010) 785; E. Manrique, M. Reuter and F. Saueressig, Annals Phys. 326 (2011) 440 and 463.
%


%Machado:2009ph
\bibitem{crehroberto}
P.F.Machado and R.Percacci, Phys.\ Rev.\ D {\bf  80} (2009)024020 and \mbox{arXiv:0904.2510 [hep-th]}.


%Bonanno:2004pq
\bibitem{bonlac}
A. Bonanno, G. Lacagnina, Nucl. Phys. B {693} (2004) 36.



%--------------------------------------------------------------------------------------------------------
\bibitem{hrt}
A. Parola, D. Pini and L. Reatto, Phys.\ Rev.\ E {\bf 48} (1993) 3321;
A. Parola and L. Reatto, Adv.\ Phys.\ 44 (1995) 211;
A. Parola, D. Pini and L. Reatto, Mol.\ Phys.\ 107 (2009) 503;
J.-M. Caillol, Nucl.\ Phys. \ B (2012) 854.



%--------------------------------------------------------------------------------------------------------
\bibitem{propref}
D. F. Litim and D. Zappal\'a, Phys. Rev. D83 (2011) 085009, \mbox{arXiv:1009.1948 [hep-th]};
P. Castorina, M. Mazza and D. Zappal\'a, Phys.\ Lett.\ B 641 (2006) 368;
M. Mazza and D. Zappal\'a, Phys.\ Rev.\ D {\bf 64} (2001) 105013;
A. Bonanno  and D. Zappal\'a, Phys. Lett. B 504 (2001) 181;
S. B. Liao, Phys. Rev. D {\bf 53} (1996) 2020.
%


%O'Raifeartaigh:1986hi
\bibitem{ora}
L. OÕRaifeartaigh, A. Wipf, H. Yoneyama, Nucl.\ Phys.\ B 271 (1986), 273.


%Wetterich:1989xg
\bibitem{wette91}
C.~Wetterich, Nucl.\ Phys.\ B 352 (1991) 529.


%Bonanno:1995ec
\bibitem{boneu}
A. Bonanno, Phys. \ Rev. D {\bf 52} (1995) 969.


%Wetterich:1992yh
\bibitem{avact}
C.~Wetterich, Phys.\ Lett.\ B 301 (1993) 90.


%--------------------------------------------------------------------------------------------------------
\bibitem{ergprop}
D. F. Litim and J. M. Pawlowski, Phys.\ Lett.\ B 546 (2002) 279;  D. F. Litim and J. M. Pawlowski, Phys.\ Lett.\ B 516 (2001) 197;


%Floreanini:1995aj
\bibitem{Flore}
R. Floreanini, R. Percacci, Phys. Lett. B356 (1995) 205. 


%Litim:2001up
\bibitem{optim}
D. F. Litim, Phys.\ Rev.\ D {\bf 64} (2001) 105007.


%Jackiw:2005yc
\bibitem{jackiw}
R.~Jackiw, C.~N\'u\~nez and S.-Y.~Pi, Phys.\ Lett.\ A 347 (2005) 47.


%
\bibitem{zakh}
E.\, C.\, Zachmanoglou and D. W. Thoe in
\textit{Introduction to Partial Differential Equations With Applications}, Courier Dover Publications, 1986.


%
\bibitem{ames}
W.\,Ames in \textit{Numerical Methods for Partial Differential Equations, Third Edition}, Academic Press, 1992.


%
\bibitem{vander}
B. Van der Pol, Edinburgh and Dublin Phil. Mag. J. of Sci., 2 (7) (1927), 978.


% Harst:2012ni
\bibitem{triat}
U.\, Harst, M.\, Reuter, \mbox{arXiv:1203.2158 [hep-th].}


%Litim:2012vz
\bibitem{cycle3}
D.\, Litim, A.\, Satz, \mbox{arXiv:1205.4218v1 [hep-th].}


%
\bibitem{priv}
D.\, Litim, A.\, Satz, private communication.


%Wilson:1970ag
\bibitem{wilson71}
K.G.~Wilson, Phys.\ Rev.\ D {bf 3} (1971) 1818.


%Bonanno:2012jy
\bibitem{bonanno12}
A.~Bonanno,  Phys.\ Rev.\ D {bf 85} (2012) 081503 and \mbox{arXiv:1203.1962 [hep-th].}


%
\bibitem{polyakov}
A.M.~Polyakov, Princeton preprint PUPTÐ1394 and \mbox{hepÐth/9304146}.


%Lauscher:2000ux
\bibitem{oliverunstable}
O.~Lauscher, M.~Reuter and C.~Wetterich, Phys.\ Rev.\  D {\bf 62} (2000) 125021 and \mbox{ hep-th/0006099 }.



%Rechenberger:2012pm
\bibitem{saur2}
S. Rechenberger, F. Saueressig,\\ \mbox{arXiv:1206.0657v1 [hep-th].}



%Benedetti:2012dx
\bibitem{dario}
D.\, Benedetti, F.\, Caravelli, JHEP 1206 (2012) 017.
%



%
%\bibitem{rafa}
%L.\, O'Raifeartaigh, A.\, Wipp, H.\, Yoneyama, Nuc. Phys, B271 (1986), 653.
%
%\bibitem{propref}
%D. F. Litim and D. Zappal\'a, Phys. Rev. D83 (2011) 085009, \mbox{arXiv:1009.1948 [hep-th]};
%P. Castorina, M. Mazza and D. Zappal\'a, Phys.\ Lett.\ B 641 (2006) 368;
%M. Mazza and D. Zappal\'a, Phys.\ Rev.\ D {\bf 64} (2001) 105013;
%A. Bonanno  and D. Zappal\'a, Phys. Lett. B 504 (2001) 181;
%S. B. Liao, Phys. Rev. D {\bf 53} (1996) 2020.
%
%\bibitem{jackiw}
%R.~Jackiw, C.~N\'u\~nez and S.-Y.~Pi,
%Phys.\ Lett.\ A 347 (2005) 47.
%
%\bibitem{hamber}
%H.W.~Hamber, Gen.\ Rel.\ Grav.\  {\bf 41} (2009) 817 and \mbox{arXiv:0901.0964 [gr-qc]};
%Phys.\ Rev.\ D 45 (1992) 507;
%Phys.\ Rev.\ D 61 (2000) 124008; \mbox{arXiv:0704.2895 [hep-th];}\\
%T.~Regge and R.M.~Williams, J.Math.Phys.41 (2000) 3964 and \mbox{gr-qc/0012035.}
%
%\bibitem{ajl1}
%J.~Ambj\o{}rn, J.~Jurkiewicz and R.~Loll,
%Phys.\ Rev.\ Lett.93 (2004) 131301.
%
%\bibitem{ajl2}
%J.~Ambj\o{}rn, J.~Jurkiewicz and R.~Loll,
%Phys.\ Lett.B 607 (2005) 205.
%%
%\bibitem{ajl3}
%J.~Ambj\o{}rn, J.~Jurkiewicz and R.~Loll,
%Phys.\ Rev.\ Lett.95 (2005) 171301;\\
%Phys.\ Rev.\ D 72 (2005) 064014;
%Contemp.Phys.47 (2006) 103.
%%
%\bibitem{back}
%L.F.~Abbott,
%Nucl.\ Phys.\ B 185 (1981) 189; Acta Phys.\ Polon.\  B {\bf 13} (1982) 33;\\
%B.S.~DeWitt, 
%Phys.Rev.162 (1967) 1195;\\
%M.T.~Grisaru, P.van Nieuwenhuizen and C.C.~Wu,
%Phys.\ Rev.\ D 12 (1975) 3203;\\
%D.M.~Capper, J.J.~Dulwich and M.\ Ramon Medrano,
%Nucl.\ Phys.\ B 254 (1985) 737;\\
%S.L.~Adler, 
%Rev.Mod.Phys.54 (1982) 729.
%%
%\bibitem{bh}
%A.~Bonanno and M.~Reuter, 
%Phys.\ Rev.\ D 62 (2000) 043008 and \mbox{hep-th/0002196;}
%Phys.\ Rev.\ D 73 (2006) 083005 and \mbox{hep-th/0602159;}\\
%Phys.\ Rev.\ D 60 (1999) 084011 and \mbox{gr-qc/9811026.}
%%
%\bibitem{erick1}
%M.~Reuter and E.~Tuiran, 
% in {\it Proceedings of the Eleventh Marcel Grossmann Meeting}, H.Kleinert, R. Jantzen, R. Ruffini (Eds.),
%World Scientific, Singapore (2007) and
%\mbox{{hep-th/0612037}}.
%%
%\bibitem{cosmo1}
%A.~Bonanno and M.~Reuter,
%Phys.\ Rev.\ D 65 (2002) 043508 and \mbox{hep-th/0106133};\\
%M.~Reuter and F.~Saueressig, 
%JCAP 09 (2005) 012 and \mbox{hep-th/0507167.}
%%
%\bibitem{cosmo2}
%A.~Bonanno and M.~Reuter,
%Phys.\ Lett.\ B 527 (2002) 9 and \mbox{astro-ph/0106468;} \\
%Int.\ J.\ Mod.\ Phys.\ D 13 (2004) 107 and \mbox{astro-ph/0210472;}\\
%E.~Bentivegna, A.~Bonanno and M.~Reuter,
%JCAP 01 (2004) 001 \\and \mbox{astro-ph/0303150.}
%%
%\bibitem{entropy}
%A.~Bonanno and M.~Reuter,
%JCAP 08 (2007) 024 and \mbox{arXiv:0706.0174 [hep-th]}; \\
%J.\ Phys.\ Conf.\ Ser.\  { 140} (2008) 012008
% and  \mbox{arXiv:0803.2546 [astro-ph]}.
%%
%\bibitem{esposito}
%A.~Bonanno, G.~Esposito and C.~Rubano,
%Gen.\ Rel.\ Grav.\ 35 (2003) 1899;\\
%Class.\ Quant.\ Grav.\ 21 (2004) 5005;\\
%A.~Bonanno, G.~Esposito, C.~Rubano and P.~Scudellaro,\\
%Class.\ Quant.\ Grav.\ 23 (2006) 3103 and 24 (2007) 1443.
%%
%\bibitem{h1}
%M.~Reuter and H.~Weyer, 
%Phys.\ Rev.\ D 69 (2004) 104022
%and \mbox{hep-th/0311196.}
%%
%\bibitem{h2}
%M.~Reuter and H.~Weyer,
%Phys.\ Rev.\ D 70 (2004) 124028
%and \mbox{hep-th/0410117.}
%%
%\bibitem{h3}
%M.~Reuter and H.~Weyer,
%JCAP 12 (2004) 001 and \mbox{hep-th/0410119.}
%%
%%
%\bibitem{girelli}
%F.~Girelli, S.~Liberati, R.~Percacci and C.~Rahmede,\\
%Class.\ Quant.\ Grav.24 (2007) 3995.
%%
%\bibitem{lhc}
%D.~Litim and T.~Plehn,
%Phys.\ Rev.\ Lett.100 (2008)131301.
%%
%\bibitem{mof}
%J.~Moffat,
%JCAP 05 (2005) 2003; \\
%J.R.~Brownstein and J.~Moffat, 
%Astrophys.\ J.\ 636 (2006) 721;\\
%Mon.\ Not.\ Roy.Astron.\ Soc.\ 367 (2006) 527.
%%
%%
%\bibitem{floper}
%R.~Floreanini and R.~Percacci,
%Nucl.\ Phys.\ B 436 (1995) 141;\\
%Phys.\ Rev.\ D 46 (1992) 1566.
%%
%\bibitem{oliverunstable}
%O.~Lauscher, M.~Reuter and C.~Wetterich,
%%\emph{Rotation symmetry breaking condensate in a scalar theory},
%  Phys.\ Rev.\  D {\bf 62} (2000) 125021 and
% \mbox{ hep-th/0006099 }.
%%
%%
%\bibitem{opt}
%D.~Litim,
%Phys.\ Lett.B 486 (2000) 92;
%Phys.\ Rev.\ D 64 (2001) 105007;\\
%Int.J.\ Mod.\ Phys.A 16 (2001) 2081.
%%
%\bibitem{giesymfp}
%H.~Gies,
%Phys.\ Rev.\ D 68 (2003) 085015.
%%
%\bibitem{nonlinsig}
%A.~Codello and R.~Percacci,
%\mbox{arXiv:0810.0715 [hep-th]}.
%%
%\bibitem{giulini}
%D.~Giulini in {\it Approaches to Fundamental Physics}, I.-O.~Stamatescu and E.~Seiler (Eds.),
%Springer, Berlin (2007).
%%
%%
%\bibitem{straumann}
%N.~Straumann in {\it Approaches to Fundamental Physics}, I.-O.~ Stamatescu and E.~Seiler (Eds.),
%Springer, Berlin (2007).
%%
%\bibitem{coscon} 
%%%%S.~Weinberg, Rev.\ Mod.\ Phys.\ {\bf 61} (1989).

\end{thebibliography}
\end{document}